\RequirePackage{fix-cm}
\documentclass[twocolumn,epjc3]{svjour3}  
\smartqed  
\RequirePackage{graphicx}
\RequirePackage[numbers,sort&compress]{natbib}
\RequirePackage[colorlinks,citecolor=blue,urlcolor=blue,linkcolor=blue]{hyperref}
\usepackage{cuted}
\usepackage{amsmath}
\usepackage{txfonts}
\usepackage{accents}
\usepackage{yfonts}

\def\lsim{\raise0.3ex\hbox{$<$\kern-0.75em\raise-1.1ex\hbox{$\sim$}}}
\def\gsim{\raise0.3ex\hbox{$>$\kern-0.75em\raise-1.1ex\hbox{$\sim$}}}

\newcommand{\be}{\begin{equation}}
\newcommand{\ee}{\end{equation}}

\newcommand{\fpmf}{f^{\pm}_f}
\newcommand{\fpmfo}{f^\pm_{0f}}
\newcommand{\fpfo}{f^+_{0f}}
\newcommand{\fmfo}{f^-_{0f}}
\newcommand{\dfpmf}{\delta f^{\pm}_f}

\newcommand{\Qpmf}{Q^{\pm}_f}

\newcommand{\apmf}{\alpha^{\pm}_f}
\newcommand{\xpmf}{x^{\pm}_f}

\newcommand{\ff}{{f'}}

\journalname{Eur. Phys. J. C}

\begin{document}

\title{Electric conductivity and flavor diffusion in a viscous, resistive quark-gluon plasma for weak and strong magnetic fields}


\author{Ferdinando Frasc\`{a}\thanksref{e1,addr1}
        \and
        Andrea Beraudo\thanksref{e2,addr1} 
        \and
        and Luca Del Zanna \thanksref{e3,addr2,addr3, addr4} 
}

\thankstext{e1}{e-mail: ferdinando.frasca@to.infn.it}
\thankstext{e2}{e-mail: andrea.beraudo@to.infn.it}
\thankstext{e3}{e-mail: luca.delzanna@unifi.it}


\institute{INFN - Sezione di Torino, Via P. Giuria 1, I-10125 Torino, Italy \label{addr1}
\and
Dipartimento di Fisica e Astronomia, Universit\`a degli Studi di Firenze, 
Via G. Sansone 1, I-50019 Sesto Fiorentino, Italy \label{addr2}
\and
INFN - Sezione di Firenze, Via G. Sansone 1, I-50019 Sesto Fiorentino, Italy \label{addr3}
\and
INAF - Osservatorio Astrofisico di Arcetri, Largo E. Fermi 5, I-50125 Firenze, Italy \label{addr4}
}

\date{Received: date / Accepted: date}

\maketitle
\begin{abstract}
We present a microscopic calculation of the electric conductivity and net-particle diffusion coefficients for a viscous and resistive ultra-relativistic plasma. 
Our results might be of interest for several astrophysical and cosmological problems, but
the main physical application we have in mind is the hot deconfined matter produced in relativistic heavy-ion collisions. Accordingly, as charged particles of the medium we take three species (flavors) of light (massless for the sake of simplicity) quarks -- $u$,  $d$ and $s$ -- and antiquarks, entailing the existence of three macroscopic conserved charges: baryon number ${\cal B}$, electric charge $Q$ and strangeness $S$.
Our results are valid both in a weakly and in a strongly-magnetized plasma, where the energy stored in the magnetic field is comparable to the one carried by the medium particles. Actually, for a conformal fluid, the behavior of the system only depends on the ratio between the thermal and the magnetic pressure, the so-called plasma beta-parameter, acting as a scaling variable.
Our calculation, starting from a relativistic Boltzmann-Vlasov equation, is based on a generalized Chapman-Enskog approach in which space-time gradients and the local electric field are treated as first-order quantities in a perturbative expansion, while terms containing magnetic corrections are considered of zeroth order and hence self-consistently resummed.
We find that, also in the strong-field limit, for each conserved charge a generalized Wiedemann-Franz law, connecting charge conductivity and diffusion coefficient, exists. However these transport coefficients acquire a non-trivial tensor structure, reflecting the development of a longitudinal, a transverse and a Hall current as a response to electric fields or density gradients.

\end{abstract}

\section{Introduction}
\label{intro}
Relativistic heavy-ion collision (HIC) experiments have been carried out for decades with the purpose of exploring the deconfined phase of the matter governed by strong nuclear interactions (QCD), the so called Quark-Gluon Plasma (QGP). For an overview of the major experimental results and of the underlying theoretical picture we refer the reader to Refs.~\cite{STAR:2005gfr,ALICE:2022wpn,Busza:2018rrf}. In HIC's, at the highest center-of-mass energies, the electric current of the protons from the colliding nuclei acts as the external source of a magnetic field $\vec B$ aligned, on average, orthogonally to the reaction plane. In the first instants after the collisions such magnetic fields are huge, even stronger than the ones in magnetars and in the early universe, reaching values of order $10^{15}-10^{16}$ T at RHIC and LHC~\cite{Skokov:2009qp,Deng:2012pc,Tuchin:2013apa}, possibly leading to the discovery of novel phenomena related to non-perturbative aspects of QCD (for a seminal paper see Ref.~\cite{Kharzeev:2007jp}) like the so-called Chiral Magnetic Effect (CME). As an independent experimental signature of these huge initial magnetic fields, the azimuthal distributions of opposite-charge particles produced in the early stages of the collision (charm and anticharm quarks of lepton-antilepton pairs from $Z^0$ boson decays) was also proposed~\cite{Das:2016cwd,Sun:2020wkg}. At the moment, experimental evidence of the CME was found in the area of condensed matter, as shown in Refs.~\cite{Xiong_2015,Li:2014bha,Huang_2015}, while its occurrence in the context of HIC's still remains elusive~\cite{ALICE:2012nhw,STAR:2014uiw}.

Clearly, the phenomenological relevance of this huge initial magnetic field and, in particular, the possibility of finding evidence of such an exotic phenomenon like the CME strongly depend on the subsequent evolution of the electromagnetic field inside the hot, deconfined fireball formed in the collisions. While in vacuum the initial magnetic field would very rapidly decay, in the opposite limit of an ideal conductor its magnitude would remain significant during the entire lifetime of the fireball, due to magnetic-flux freezing. Hence, the possible impact of electromagnetic fields on final-state observables is strongly related to the actual value of the electric conductivity of the QGP.

In order to include the above effects into a realistic modeling of the fireball one has to supplement current relativistic hydrodynamics codes (see Refs.~\cite{Rocha:2023ilf,Shen:2020mgh,Gale:2013da} for a review) with further equations describing the self-consistent evolution of the electromagnetic field coupled to the charged particles of the medium. This can be done within the framework of relativistic magneto-hydrodynamics (MHD), a standard tool originally developed to address astrophysical problems, where often general relativity (GR) is also needed when dealing with the extreme gravity by compact objects. Examples of application of relativistic MHD and GRMHD models to astrophysical problems range from magnetized neutron stars~\cite{Komissarov:2005xc,Pili:2014npa}, to pulsar wind nebulae~\cite{Komissarov:2004,DelZanna:2006wj}, to extragalactic jets~\cite{Mignone:2010}, to accretion onto black holes~\cite{McKinney:2004ka,EventHorizonTelescope:2019pcy}. Relativistic MHD equations are often solved in the limit of an ideal electric conductor. However, as pointed out in Ref.~\cite{Denicol:2019iyh}, the electric conductivity $\sigma_Q$, as the other transport coefficients (shear viscosity, charge diffusion, etc.), is proportional to the particle mean-free path: hence, taking the $\sigma_Q\to\infty$ limit while keeping the rest of the transport coefficients constant (and small, if a fluid description has to be applied) is not consistent from a conceptual point of view. Notice, by the way, that an important phenomenon occurring in many astrophysical plasmas -- magnetic reconnection -- can only take place in the presence of an effective resistivity of the medium, even in the relativistic regime~\cite{DelZanna:2016uiq,Bugli:2024fby}, and resistive simulations of relativistic astrophysical sources are a current field of research \cite{Tomei:2019zpj,Ripperda:2021zpn,Nathanail:2021jbn,Mattia:2024erd}.

Formulating a resistive MHD description of the matter produced in HIC's, in which all possible dissipative effects (viscous friction, particle diffusion and Joule-heating) contributing to entropy production are taken into account, is a very ambitious task.  This challenge was addressed for instance in Refs.~\cite{Denicol:2019iyh,Panda:2021pvq,Kushwah:2024zgd} within a kinetic description. 
A more macroscopic approach to obtain analogous results would consist in the  generalization of the usual M\"uller-Israel-Stewart (MIS) theory~\cite{Mueller1967,Israel:1976tn,Israel:1979wp}, based on the imposition of a local non-negative rate of entropy production. In both cases the resulting equations for the dissipative fields contain a set of transport coefficients, governing the diffusion of conserved charges: energy, momentum, baryon number, electric charge and strangeness for the case of 3 quark flavors. Fixing the latter in the presence of strong electromagnetic fields, when the magnetic and thermal pressures are comparable, is a highly non-trivial problem, due to the breaking of isotropy and the appearance of new dimensionful scales which complicate the usual power-counting schemes adopted to get a closed system of equations.
This is the challenge which we plan to address in the rest of this paper. In particular, we focus on the problem of quark-flavor diffusion and conductivity, since -- as already occurring in the weak-field case -- the same transport coefficients (acquiring a tensor structure in the strong B-field regime) describe the response of the medium to density gradients or electric fields. Depending on the circumstances it may result more convenient to implement conservation laws either for the net-quark numbers ($u$, $d$ and $s$) or for the macroscopic charges (baryon number $\cal{B}$, electric charge $Q$ and strangeness $S$), the corresponding currents being connected by a simple change of basis.

Concerning our approach, our starting point is the  relativistic Boltzmann-Vlasov equation for the quark (and antiquark) phase-space distribution, written in relaxation-time approximation. We then perform a generalized  Chapman-Enskog expansion, in which off-equilibrium corrections are truncated at first-order in the space-time gradients and in the electric field (the local electric field vanishing in the limit of an ideal conductor), but are self-consistently resummed to all orders in the magnetic field. For the sake of simplicity we consider a conformal plasma. In this case one can conveniently rescale all dimensionful quantities, including the local magnetic field, by proper powers of the temperature. The resulting dimensionless quark-diffusion/conductivity tensor is then a function of a simple scaling variable: the so-called plasma beta-value $\beta_V$, i.e. the ratio between the thermal and magnetic pressure. The latter is the quantity determining whether the system is either weakly or strongly magnetized. In a MHD-description of the medium produced in HIC's the condition $\beta_V^{-1}\ll 1$ holds for most of the fluid cells. However, at very early times and in the most peripheral regions of the fireball  $\beta_V^{-1}$ can become of order 1 or higher. Hence, knowing the transport properties of the medium also in the strongly-magnetized regime is crucial. Notice that in astrophysical applications of the MHD equations situations in which $\beta_V^{-1}$ is not small at all are quite common, from the solar atmosphere~\cite{Bourdin_2017} to the pulsar magnetosphere~\cite{Belyaev:2014fla}, so that our analysis may result of even stronger interest for the study of these systems. Due to the nature of the Lorentz force we find that, for a strongly-magnetized QGP, quark diffusion/conductivity along the magnetic-field direction is unaffected, but in the plane orthogonal to the latter is suppressed. Furthermore, for non-zero quark chemical potential, a Hall current also develops in response to a density gradient or to an electric field.

In our paper natural units are employed, where the following universal constants are taken to be unitary, i.e. $k_B \equiv 1$ (Boltzmann constant), $c \equiv 1$ (speed of light), $\hbar \equiv 1$ (reduced Planck constant). {All electric charges are measured in units of $|e|$, with $e$ being the elementary electron charge}, such that the fine-structure constant is given by $\alpha_{\rm em}  \equiv e^2/4 \pi = 1/137$. Moreover, in this work we will adopt the (-, +, +, +) convention for the signature of the metric tensor $g_{\mu\nu}$.

Throughout the paper we will use the following compact notations 
\begin{equation}
    A_{(\mu\nu)} \equiv \textstyle{\frac{1}{2}} \left(A_{\mu\nu} + A_{\nu\mu} \right) \quad{\rm and} \quad A_{[\mu\nu]} \equiv \textstyle{\frac{1}{2}} \left(A_{\mu\nu} - A_{\nu\mu} \right) 
\end{equation}
to symmetrize/anti-symmetrize an arbitrary rank-2 tensor.
Furthermore, after introducing the projector
\begin{equation}
\Delta^{\mu\nu} \equiv g^{\mu\nu} + u^\mu u^\nu \,
\end{equation}
onto the hypersurface normal to an arbitrary time-like versor $u^\mu$ (hence $u^\mu u_\mu = -1$ and $\Delta^{\mu\nu} u_\nu = 0$), later on identified with the fluid four-velocity, we also employ the notation
\begin{equation}
    A_{\langle \mu\nu \rangle} \equiv \Delta^{\lambda \rho}_{\,\,\,\,\,\,\,\mu \nu} \, A_{\lambda\rho} 
\end{equation}
to indicate its symmetric, transverse and traceless part, where
\begin{equation}
    \Delta^{\mu \nu}_{\,\,\,\,\,\,\,\lambda \rho} \equiv \frac{1}{2} \left( \Delta^\mu_{\,\,\,\,\lambda} \, \Delta^\nu_{\,\,\,\,\rho} +  \Delta^\mu_{\,\,\,\,\rho} \, \Delta^\nu_{\,\,\,\,\lambda}\right) - \frac{1}{3} \Delta^{\mu \nu} \Delta_{\lambda\rho} \,.
\end{equation}
Since more convenient sets of coordinates, rather than the Minkowskian one, are often employed in the study of HIC's\footnote{{e.g. the Bjorken coordinates $(\tau,x,y,\eta_s)$, in which $\tau\!\equiv\!\sqrt{t^2-z^2}$ and $\eta_s\!\equiv\!{\rm atanh}(z/t)$}, giving a non-trivial metric tensor $g_{\mu\nu}\!=\!\rm{diag}(-1,1,1,\tau^2)$}, with non-vanishing Christoffel symbols~\cite{Bjorken:1982qr,Romatschke:2009im}, it looks natural (and suited to a direct generalization to astrophysical problems) to directly provide a description of our system within the framework of GRMHD. 
Hence, from now on the metric tensor will be a generic $g_{\mu\nu}$ rather than the Minkowskian $\eta_{\mu\nu}$, and we will use (geometric) covariant derivatives $\nabla_\mu$, whose action on a generic vector/tensor is given by~\cite{Carroll_2019,Misner:1973prb,Weinberg:1972kfs}
\begin{align}
\label{eq:covder1}
& \nabla_\mu \,A_{\alpha}  \equiv \partial_\mu \,A_{\alpha} -
\Gamma^{\lambda}_{\,\,\,\,\mu\alpha} \, A_{\lambda}\,\,\\ 
& \nabla_\mu \,B^{\alpha\beta}  \equiv \partial_\mu \, B^{\alpha\beta} + 
\Gamma^{\alpha}_{\,\,\,\,\mu\lambda} \, B^{\lambda\beta}+
\Gamma^{\beta}_{\,\,\,\,\mu\lambda} \, B^{\alpha\lambda}\,, \label{eq:covder2}
\end{align}
where the Christoffel symbols read
\begin{equation}
\label{eq:christoffel}
\Gamma^{\lambda}_{\,\,\,\,\alpha\beta}  =\tfrac{1}{2} g^{\lambda\gamma} 
\left(\partial_{\alpha} \, g_{\gamma\beta} + \partial_{\beta} \, g_{\alpha\gamma}
- \partial_{\gamma} \, g_{\alpha\beta} \right).
\end{equation}
In addition, we define
\begin{equation}
D \equiv u^{\mu} \, \nabla_{\mu}
\label{eq:comoving}
\end{equation}
as the comoving (covariant) derivative, according to the decomposition~\cite{Song:2007ux,Romatschke:2009im}
\be
\nabla_\mu\equiv -u_\mu \, D+\Delta_\mu\,,\quad{\rm with}\quad \Delta_\mu\equiv \Delta_{\mu\nu}\nabla^\nu\,,
\ee
where the last term is the derivative normal to the velocity $u^\mu$.
We also introduce the general decomposition~\cite{Muronga:2003ta,Baier:2007ix,Romatschke:2009im}
\begin{equation}
\Delta_\mu u_\nu = \sigma_{\mu\nu} + \omega_{\mu\nu} +  \frac{1}{3} \Delta_{\mu\nu} \, \Theta\,,
\label{eq:identity}
\end{equation}
where
\begin{equation}
\Theta\equiv\nabla_\mu \, u^\mu = \Delta_\mu \,\label{eq:exp-rate} u^\mu
\end{equation}
is the expansion rate of the fluid, 
\begin{equation}
    \sigma_{\mu\nu} \equiv \Delta_{\langle \mu} u_{\nu \rangle} = \Delta_\mu^{\,\,\,\,\lambda} \Delta_\nu^{\,\,\,\,\rho}
\left[\textstyle{\frac{1}{2}}\,\bigl(\nabla_\lambda u_\rho + \nabla_\rho u_\lambda \bigr)
- \textstyle{\frac{1}{3}}\Delta_{\lambda\rho}\,\Theta \right] \label{eq:shear}
\end{equation}
is the shear tensor, that is a symmetric, traceless and transverse tensor based on first-order derivatives of $u^\mu$, and finally
\begin{equation}
\omega_{\mu\nu} \equiv \Delta_{[\mu} \, u_{\nu]}   = \Delta_\mu^{\,\,\,\,\lambda} \Delta_\nu^{\,\,\,\,\rho}
\left[\textstyle{\frac{1}{2}}\, \bigl(\nabla_\lambda u_\rho - \nabla_\rho u_\lambda \bigr) \right]  
\label{eq:vorticity}
\end{equation}
represents the vorticity tensor of the fluid, an antisymmetric, traceless and transverse tensor.

\section{GRMHD equations for multiple conserved charges}\label{sec:GRMHD}
In studying the strongly-interacting QCD plasma formed in HIC's, GRMHD provides an effective long-wavelength description of the evolution of the bulk produced matter including in a self-consistent way -- beside nuclear interactions -- the coupling of electrically charged particles with the electromagnetic field.

GRMHD is based on the numerical implementation of a set of conservation laws for the $N_f$ net-quark currents (flavor-changing weak interactions occur on much longer time-scales than the fireball lifetime) and the total energy-momentum tensor
\begin{align}
& \nabla_\mu J_f^\mu  =  0  \,,\label{eq:curr}\\
& \nabla_\mu T^{\mu\nu}  =  (J_{\rm ext})_\mu \, F^{\mu\nu} \label{eq:enmom} \,.
\end{align}
In the above $f=u,d,s$ (in HIC's charm and bottom quarks are external probes which do not have enough time to reach thermodynamic equilibrium and top quarks decay before the ``hydrodynamization'' of the system\footnote{We refer the reader to Refs.~\cite{Soloviev:2021lhs,Strickland:2024moq} for more information about the occurrence of a collective behavior in the medium produced in HIC's and the related emergence of hydrodynamic attractors in kinetic theory}).
In the second equation $T^{\mu\nu}$ includes both a matter and an electromagnetic contribution, namely:
\begin{equation}
    T^{\mu\nu}\equiv T^{\mu\nu}_{\rm m}+T^{\mu\nu}_{\rm em}\,,
\end{equation}
which will be later specified. Moreover, $J_{\rm ext}^\mu$ represents an external electric current, in practice the spectator protons of the colliding nuclei, acting as a source of the initial huge magnetic field. Since they leave the interaction region at approximately the speed of light, it will not be relevant for the rest of our discussion.
The conserved net-quark currents arise from the difference of the particle (+) and antiparticle (-) contributions, namely 
\begin{equation}
J_f^\mu\equiv J^{+\mu}_f-J^{-\mu}_{f}\,.\label{eq:netquarkdef}
\end{equation}
Since in HIC's, even at the highest center-of-mass energies, after about 10 fm/c hadronization occurs, it is more convenient to identify three ``macroscopic'' conserved charges, which can be measured starting from the yields of the final detected hadrons: the baryon number $\cal B$, the electric charge $Q$ (responsible for the coupling with the electromagnetic field) and the strangeness $S$. Within the framework of Quantum Field Theory these quantities are defined as
\begin{eqnarray}
\cal B &\equiv&\int d^3 x \,\,\psi^\dagger(x)\,\varmathbb{B}\,\psi(x)\,,\\
Q &\equiv& |e|\!\!\int d^3 x\,\, \psi^\dagger(x)\,\varmathbb{Q}\,\psi(x)\,,\\
S &\equiv&\int d^3 x\,\, \psi^\dagger(x)\,\varmathbb{S}\,\psi(x)\,,
\end{eqnarray}
where $\psi^{\rm T} \equiv (\psi_u, \psi_d, \psi_s)$ is an array containing the spin-1/2 quark fields and 
\begin{eqnarray}
\varmathbb{B} &\equiv& {\rm diag}(\varmathbb{B}_u,\varmathbb{B}_d,\varmathbb{B}_s)=(1/3,1/3,1/3) \,,\\
\varmathbb{Q} &\equiv& {\rm diag}(\varmathbb{Q}_u,\varmathbb{Q}_d,\varmathbb{Q}_s)=(2/3,-1/3,-1/3)\,,\\
\varmathbb{S} &\equiv& {\rm diag}(\varmathbb{S}_u,\varmathbb{S}_d,\varmathbb{S}_s)=(0,0,-1)\,,
\end{eqnarray}
are the operators associated with the conserved charges.
It is then possible to express the ``macroscopic'' currents in terms of the quark-flavor ones, defined in Eq.~(\ref{eq:netquarkdef}), as follows:
\begin{equation}
J_B^\mu\equiv\sum_f^{u, d, s} \varmathbb{B}_f \, J_f^\mu\,,\quad
J_Q^\mu\equiv\sum_f^{u, d, s} \varmathbb{Q}_f \, |e| \, J_f^\mu\,,\quad
J_S^\mu\equiv\sum_f^{u, d, s} \varmathbb{S}_f \, J_f^\mu\,,\quad\label{eq:conserved-currents}
\end{equation}
which must satisfy the continuity equations $\nabla_\mu \, J_q^\mu \equiv 0$, where $q \in \{{\cal B}, Q, S \}$. One can rephrase the equalities in Eq.~(\ref{eq:conserved-currents}) in a more compact form, i.e.
\begin{equation}
    J_q^\mu\equiv{\cal M}_{qf}\,J_f^\mu\quad{\rm and}\quad J_f^\mu\equiv \left({\cal M}^{-1}\right)_{fq}J_q^\mu\,,\label{eq:qfbasis1}
\end{equation}
where the following matrices
\begin{equation}
\label{eq:M}
{\cal M}\equiv
\begin{pmatrix}
    \frac{1}{3} & \frac{1}{3} & \frac{1}{3}\\
    \frac{2}{3}|e| \,&\, -\frac{1}{3}|e|\, &\, -\frac{1}{3}|e| \\
    0 & 0 & -1
\end{pmatrix}
\quad{\rm and}\quad
{\cal M}^{-1}\equiv
\begin{pmatrix}
    1 & \frac{1}{|e|} & 0\\
    2\, &\, -\frac{1}{|e|}\, &\, 1\\
    0 & 0 & -1
\end{pmatrix}
\end{equation}
have been introduced.
{Thermodynamic relations must be generalized to the case of multiple conserved charges, each one described by an independent chemical potential which we can collect into the vectors
\begin{equation}
    \mu_q^{\rm T}\equiv(\mu_{\cal B},\mu_Q,\mu_S)\,,\quad \mu_f^{\rm T}\equiv(\mu_{u},\mu_d,\mu_d)\,,
\end{equation}
depending whether one is working in the charge or flavor basis, connected by
\begin{equation}
\mu_f=\varmathbb{B}_f\,\mu_{\cal B}+|e|\,\varmathbb{Q}_f\,\mu_Q+\varmathbb{S}_f\,\mu_S\equiv \left({\cal M}^{\rm T}\right)_{fq}\,\mu_q\,,\label{eq:muf-muq}
\end{equation}
ensuring that
\begin{equation}
\mu_q^{\rm T} \, n_q=\mu_q^{\rm T}\,{\cal M}_{qf}\,n_f\equiv \mu_f^{\rm T}\,n_f\,.    
\end{equation}
The Euler relation} then reads:
\begin{equation}
    \varepsilon + P = T s + \sum_q^{{\cal B},Q,S} \mu_q^{\rm T} \, n_q= T s + \sum_f^{u,d,s}\mu_f^{\rm T} \, n_f \,,\label{eq:Euler-multiple}
\end{equation}
{ while for} the first principle of thermodynamics {one has}
\begin{equation}
    \mathrm{d} \varepsilon = T\mathrm{d}s + \sum_q^{{\cal B},Q,S} \mu_q^{\rm T} \, \mathrm{d}n_q=
T\mathrm{d}s + \sum_f^{u,d,s} \mu_f^{\rm T} \, \mathrm{d}n_f\,.\label{eq:1stlaw-multiple}
\end{equation}
Accordingly
\begin{equation}
dP=s\,dT+ \sum_q^{{\cal B},Q,S}n_q\,d\mu_q= s\,dT+ \sum_f^{u,d,s}n_f\,d\mu_f\,.\label{eq:dP}
\end{equation}
From Eqs.~(\ref{eq:1stlaw-multiple}) and~(\ref{eq:dP}) the relations
\begin{equation}
T=\left(\frac{\partial\varepsilon }{\partial s}\right)_{n_f}\,,\quad \mu_f=\left(\frac{\partial\varepsilon }{\partial n_f}\right)_{s,n_{f'}\ne n_f}\,,
\end{equation}
allowing one to define a local temperature\footnote{In a general-relativistic context, however, one has to be careful when dealing with the concept of temperature. Thus, for a fluid in thermodynamic equilibrium, embedded in a curved stationary spacetime, $T(x)$ should satisfy a Tolman-type law, in order to prevent the violation of the principles of thermodynamics~\cite{Santiago:2018lcy,Kovtun:2022vas} -- a condition that is actually frame-dependent.} and chemical potential,
and
\begin{equation}
\label{eq:therm_rel}
s=\left(\frac{\partial P}{\partial T}\right)_{\mu_f}\,,\quad   n_f= \left(\frac{\partial P}{\partial \mu_f}\right)_{T,\,\mu_{f'}\ne\mu_f}\,,
\end{equation}
follow. Analogous expressions are obtained using the macroscopic charge basis.

In addition to the conservation laws in Eqs.~(\ref{eq:curr}) and~(\ref{eq:enmom}) one must self-consistently describe the evolution of the electromagnetic field (we refer the reader to Ref.~\cite{Inghirami:2016iru} for details), which obeys the Maxwell equations
\begin{align}
& \nabla_\mu F^{\mu\nu} = - J^\nu  \label{eq:Max1}\,, \\
& \nabla_\mu \hspace{0.01cm} ^{\star}\!F^{\mu\nu} = 0 \,,
\end{align}
where $F^{\mu\nu}$ is the Faraday tensor, whereas
$^{\star}\!F^{\mu\nu} \equiv \textstyle{\frac{1}{2}}\epsilon^{\mu\nu\lambda\rho}F_{\lambda\rho}$ represents its dual. Furthermore, $J^\nu \equiv J_{\rm ext}^\nu + J_Q^\nu$ is the total electric current, source of the electromagnetic field acting on the plasma, whose quark contribution was defined in Eq.~(\ref{eq:conserved-currents}). Notice that Eq.~(\ref{eq:Max1}) automatically ensures charge conservation in the plasma,
\begin{equation}
\nabla_\mu J^\mu_Q = 0 \,,
\end{equation}
since also the spectator protons are clearly conserved.

After defining, within the single-fluid approximation, a common four-velocity $u^\mu$ for the plasma (a special care is required in the case of a dissipative medium, as discussed in Sect.~\ref{sec:dissipative}) the Faraday tensor and its dual can be split with respect to the latter as~\cite{Anile1990,Hattori:2022hyo}
\begin{align}
F^{\mu\nu} & = u^\mu e^\nu - u^\nu e^\mu + \epsilon^{\mu\nu\lambda\kappa} b_\lambda \, u_\kappa\,, \\
^{\star}\!F^{\mu\nu} & = u^\mu b^\nu - u^\nu b^\mu - 
\epsilon^{\mu\nu\lambda\kappa} e_\lambda \, u_\kappa\,,
\end{align}
where
\begin{align}
e^\mu  & \equiv F^{\mu\nu}\, u_\nu \,,  \quad (e^\mu u_\mu =0) \,, \\
b^\mu  & \equiv \hspace{0.01cm} ^{\star}\!F^{\mu\nu} \, u_\nu \,, \quad  (b^\mu u_\mu =0)\,,
\end{align}
are the electric and magnetic fields measured in the comoving frame of the fluid.

It is noteworthy that, here, we have neglected possible polarization and magnetization
effects
{associated to the dipole moments of the plasma particles}, therefore we do not make distinction between microscopic and
macroscopic fields~\cite{Felderhof2004,Israel:1978up,Kovtun:2016lfw,Jou_1988,dixon1978}.
Under this assumption, the electromagnetic contribution to the energy-momentum tensor is known to be~\cite{Anile1990,DelZanna:2018dyb}
\be
T^{\mu\nu} _\mathrm{em} = F^{\mu\lambda}F^\nu_{\,\,\,\,\lambda} - 
\tfrac{1}{4}\,g^{\mu\nu}F^{\lambda\kappa}F_{\lambda\kappa},
\ee
whose evolution law $\nabla_\mu T^{\mu\nu} _\mathrm{em} = J_\mu\,F^{\mu\nu}$ follows from the Maxwell equations. Hence, one has for the matter contribution to the energy-momentum tensor
\be
\nabla_\mu T^{\mu\nu} _\mathrm{m}=-(J_Q)_\mu\,F^{\mu\nu}\,,\label{eq:dmuTmunu-matt}
\ee
where the RHS arises from the Lorentz force acting on the plasma particles, which entails a continuous energy transfer between the matter and field degrees of freedom.

The system of GRMHD equations must be closed by an Equation of State (EoS), in the following conveniently written in terms of the quark chemical potentials, i.e.
\begin{equation}
    P=P(T,\{\mu_f\})\,.\label{eq:EoS-def}
\end{equation}

During the system evolution the second law of thermodynamics in its local form
\be
\nabla_\mu \, \mathcal{S}^{\,\mu}\ge 0\,,\label{eq:2ndlaw}
\ee
$\mathcal S^{\,\mu}$ being the entropy four-current, must be satisfied. In the absence of dissipation the vanishing of the four-divergence of the entropy current automatically follows from the ideal GRMHD set of equations. On the other hand, if non-ideal effects (viscosity, diffusion/heat-conduction and/or resistivity) are at work, Eq.~(\ref{eq:2ndlaw}) provides a way to fix the dissipative contributions to the energy-momentum tensor and to the conserved net-particle currents. These last issues will be discussed in the following sections.

Consider now the particular case of a Minkowski-type spacetime, that is, following the notation of \cite{Inghirami:2016iru}, with a metric written as
\be
\mathrm{d}s^2 = - (\mathrm{d}x^0)^2 + g_{ij}\, \mathrm{d}x^i \mathrm{d}x^j\,,
\ee
where latin indices span spatial components from 1 to 3, and $g_{ij}$ is not necessarily diagonal and it may depend on space but also on time\footnote{Bjorken coordinates also share this form of the metric. For a rigorous generalization to curved spacetimes, we refer the reader, for instance, to Refs.~\cite{DelZanna:2007pk,Rezzolla-Zanotti2013,Chabanov:2021dee}.}. Then, the four-velocity can be written as
\be
u^\mu = \gamma \, (1, v^i )\,\,,
\ee
with $v^i$ the three-velocity of the fluid and $\gamma = 1/\sqrt{1-v_kv^k}$ the corresponding Lorentz factor.
Moreover, the electromagnetic fields can be split as~\cite{Mignone:2019ebw}
\begin{align}
e^\mu &= \gamma \, \left( v_kE^k, E^i + \epsilon^{ijk}v_jB_k \right) \,, \label{eq:Elab}\\
b^{\mu} &= \gamma \, \left( v_kB^k, B^i - \epsilon^{ijk}v_jE_k \right) \,,
\end{align}
where $E^i$ and $B^i$ are, respectively, the electric and magnetic fields in the usual LAB frame, and $\epsilon^{ijk} = |\mathrm{det}(g_{ij})|^{-1/2}[ijk]$ is the usual Levi-Civita pseudo-tensor associated to the spatial three-metric, with $[ijk]$ the alternating symbol with values $\pm 1$ or 0.

\subsection{Ideal GRMHD equations}
\label{sec:ideal}
In the ideal case both the energy-momentum tensor and the conserved currents admit a unique well defined decomposition in terms of the fluid four-velocity $u^\mu$ and the spacetime metric $g_{\mu \nu}$ only~\cite{Romatschke:2017ejr,Rezzolla-Zanotti2013}. The fluid contribution to the energy momentum tensor reads
\begin{equation}
    T^{\mu\nu}_\mathrm{m} =
\varepsilon \, u^\mu u^\nu + P \Delta^{\mu\nu}\,.
\end{equation}
In the above (zeroth-order {in a gradient expansion}) expression $\varepsilon \equiv T^{\mu\nu}_\mathrm{m}u_\mu u_\nu $ represents the {fluid} energy density and $P \equiv \tfrac{1}{3} \Delta_{\mu\nu} T^{\mu\nu}_\mathrm{m}$ its pressure.
{The currents associated with the quark flavor $f$ and the macroscopic conserved charges flow parallel to the fluid velocity (no particle diffusion occurs in the ideal case):
\be
J_f^\mu=n_f\, u^\mu\,,\quad J_q^\mu=n_q\, u^\mu\,.\label{eq:J-id}
\ee
In the above $n_f\!=\!-J^\mu_f \, u_\mu$ and $n_q\!=\!-J^\mu_q \, u_\mu$ are the net-quark and charge density measured in the comoving frame and are related by Eq.~(\ref{eq:qfbasis1}). In the fireball produced in HIC's an average non-vanishing net-quark density can be there only if some stopping of the incoming nuclear matter takes place. This  stopping is completely negligible at LHC energies, where an exact particle-antiparticle balance is found~\cite{ALICE:2023ulv}, but it is relevant in collisions performed at lower center-of-mass energies. In this last case $n_{\cal B},n_Q\ne 0$ (but $n_S=0$) and non-vanishing convective currents parallel to the fluid velocity $u^\mu$ can develop in the system. 

Concerning the electromagnetic field, the ideal behavior of the system -- corresponding to an infinite conductivity -- translates into the vanishing of the electric field in the local rest frame (LRF) of the plasma~\cite{Anile1990}, i.e.
\be
e^\mu=0\quad\longrightarrow\quad F^{\mu\nu}=\epsilon^{\mu\nu\lambda\kappa} b_\lambda \, u_\kappa \equiv b^{\mu\nu}\,,\label{eq:Eideal-cond}
\ee
so that the electromagnetic contribution to the stress-energy tensor reads~\cite{Inghirami:2016iru}
\be
T^{\mu\nu}_{\rm em}=\frac{1}{2}b^2 u^\mu u^\nu+\frac{1}{2}b^2 \Delta^{\mu\nu}-b^\mu b^\nu
\,. 
\ee
In a Minkowski-type spacetime, according to Eq.~(\ref{eq:Elab}), the vanishing of $e^\mu$ entails
\begin{equation}
E^i + \epsilon^{ijk} v_j B_k = 0\,,\label{eq:Eidealcond}
\end{equation}
meaning that the electric field (the one measured in the LAB frame) is no longer an independent variable. Moreover, the magnetic field in the LRF of the plasma is
\be
b^\mu = \left[ \gamma \,(v_k B^k), B^i/\gamma + \gamma \, (v_k B^k) \,v^i \right]\,,
\ee
and
\be
b^2 = (B_k B^k)/\gamma^2 + (v_k B^k)^2\,.
\ee

We now write the equations implementing the conservation of net-quark number, energy and linear momentum in the ideal regime. From Eq.~(\ref{eq:J-id}) one gets:
\begin{equation}
Dn_f+n_f\Theta=0\,,\quad Dn_q+n_q\Theta=0\,,\label{eq:fq-id}
\end{equation}
written in terms of the comoving derivative and of the expansion rate defined in Eqs.~(\ref{eq:comoving}) and~(\ref{eq:exp-rate}).
Taking the projection of Eq.~(\ref{eq:dmuTmunu-matt}) along the fluid four velocity, one has
\be
u_\nu\nabla_{\mu}T^{\mu\nu}_{\rm m}=0\quad\longrightarrow\quad
D\varepsilon+(\varepsilon+P)\,\Theta=0\,.\label{eq:en_id}
\ee
Notice that the RHS of Eq.~(\ref{eq:dmuTmunu-matt}) vanishes for an ideal conductor, in which
\begin{equation*}
    (J_Q)_\mu \, F^{\mu\nu}=n_Q \, u_\mu \, F^{\mu\nu} = -n_Q \, e^\nu=0\,.
\end{equation*}
{Finally, taking the  projection of Eq.~(\ref{eq:dmuTmunu-matt}) orthogonal to the fluid four-velocity one gets the well-known relativistic Euler equation for the fluid acceleration $a_{\mu} \equiv D u_{\mu}$, given by: 
\begin{equation}
\left(\varepsilon + P\right) a_{\mu} + \Delta_{\mu} P  = 0 \label{eq:mom_id} \,.
\end{equation}
} 

In an ideal plasma, entropy is exactly conserved during its evolution and consequently the entropy current
\be
\mathcal{S}^\mu \equiv s \, u^{\mu}\label{eq:scurrent-id}
\ee
obeys the continuity equation
\begin{equation}
\nabla_\mu \, \mathcal{S}^{\,\mu} = 0 \,.
\label{eq:entropy}
\end{equation}
In the above, $s \equiv -u_\mu \, \mathcal{S}^\mu$ is the entropy density in the comoving frame.
{Notice that Eq.~(\ref{eq:scurrent-id}) for the entropy current only holds in the ideal case. Its corrections appearing in a dissipative plasma will be addressed in the following sections.} Actually Eq.~(\ref{eq:entropy}) is not an independent equation, but directly stems from energy conservation. In fact, after substituting Eqs.~(\ref{eq:Euler-multiple}) and~(\ref{eq:1stlaw-multiple}) into Eq.~(\ref{eq:en_id}) and taking into account the net-quark number conservation in Eq.~(\ref{eq:fq-id}) one gets
\be
Ds+s\,\Theta=0\,,
\ee
recovering the continuity equation in Eq.~(\ref{eq:entropy}). The latter, {together with the equation for baryon-number conservation} entails that, for an ideal plasma, the entropy per baryon is a conserved quantity along each streamline, i.e.
\begin{equation}
    D(s/n_B)=0\,.
\end{equation}
\subsection{Dissipative GRMHD equations}
\label{sec:dissipative}
The description of a hot plasma of charged particles as an inviscid ideal conductor, although widely employed in the literature, is affected by a conceptual inconsistency. Considering as a guidance the predictions from kinetic theory, one gets that both the shear viscosity $\eta=\frac{1}{5}{\tau_R}\,(\varepsilon+P)$ and the electric conductivity $\sigma_{Q}\sim {\tau_R}\,e^2T^2$ are proportional to the particle relaxation time: one cannot have at the same time an inviscid fluid ($\eta\to 0$) and an ideal electric conductor ($\sigma_Q\to \infty$). 
It is then necessary to include all possible
non-ideal effects, including particle diffusion ({or heat conduction}), bulk and shear viscosities and electric resistivity. First of all, for a relativistic {dissipative} fluid, one has to establish what is the meaning of the fluid four-velocity. Typically, in the context of {high-energy} HIC's {in which the net-quark density can be very small or even vanishing}, the so-called Landau-Lifshitz frame~\cite{Landau1987Fluid} is preferred.
{In this case} $u^\mu$ corresponds to the velocity of energy transport and, accordingly, there is no heat flux from a fluid cell to an adjacent one {at a different temperature}. Consequently, all primary fluid quantities must be modified with the introduction of dissipative fluxes, namely~\cite{Romatschke:2017ejr,Song:2007ux}
\begin{align}
& J_f^\mu =  n_f \, u^\mu  + j_f^\mu \,, \label{eq:diff-f}\\
& T^{\mu\nu}_\mathrm{m} = \varepsilon \, u^\mu u^\nu + (P + \Pi) \, \Delta^{\mu\nu} + \pi^{\mu\nu} \,. \label{eq:Tm} 
\end{align}
{Notice that the quark-flavor currents $J^\mu_f$ are no longer aligned with the fluid velocity $u^\mu$, but receive a contribution $j^\mu_f\equiv\Delta^{\mu}_{\,\,\,\,\nu} J^\nu_f$ orthogonal to the latter and describing particle diffusion.

As far as the matter energy-momentum tensor is concerned {two dissipative terms appear: a bulk viscous pressure}
\begin{equation}
\Pi \equiv {\frac{1}{3}} \Delta_{\mu\nu} \, T_\mathrm{m}^{\mu\nu} -P
\end{equation}
and {the shear viscous tensor}
\begin{equation}
\pi^{\mu\nu} \equiv \Delta^{\mu \nu}_{\,\,\,\,\,\,\,\lambda \rho} \, T_\mathrm{m}^{\lambda\rho}\,,
\end{equation}
symmetric, traceless ($\pi^\mu_{\,\,\,\,\mu} =0$) and transverse to the fluid velocity ($ \pi^{\mu\nu}u_\nu =0$). As shown in the following, these non-ideal corrections to the energy-momentum tensor ($\pi^{\mu \nu}$ and $\Pi$) and {to the net-quark} currents ($j_f^\mu$) are at least of first order in the gradients
{and} must vanish as the system approaches thermodynamic equilibrium ($\pi^{\mu\nu}$, $\Pi$, $j_f^\mu$ $\rightarrow$ $0$). 

{Notice that in the Eckart frame, in which the fluid velocity is defined by the (net) particle current, one would have $j_f^\mu=0$, while the matter energy-momentum tensor would receive an extra contribution from the heat flux~\cite{Eckart_1940}:
\begin{equation*}
    T^{\mu\nu}_{\rm m}\longrightarrow T^{\mu\nu}_{\rm m}+ 2 \, h^{(\mu} u^{\nu)} \,\,,\quad{\rm with}\quad u_\mu \, h^\mu=0\,.
\end{equation*}
Since in the realistic case multiple quark flavors with possibly different transport coefficients are present in the system, the choice of the Landau frame looks more natural to us.}

The requirement of orthogonality for the non-ideal terms in Eqs.~(\ref{eq:diff-f}) and~(\ref{eq:Tm}) give rise to the so-called Landau matching conditions, which are given by
\begin{align}
\label{eq:matching1}
&\varepsilon = u_\mu u_\nu \, T^{\mu \nu}_{\rm m} = \varepsilon_{\rm eq} \,, \\
&n_f = - u_\mu \, J_f^{\mu} = n_{f, \rm {eq}}\label{eq:matching_f}
\end{align}
{and, accordingly,}
\begin{equation}
\label{eq:matching2}
n_q = - u_\mu \, J_q^{\mu} = n_{q, \rm {eq}} \,, \,\,\,\,
q\in\{{\cal{B}},Q,S\}
\,.
\end{equation}
{Physically this means that, in the local rest frame (LRF), the energy and flavor/charge densities coincide with the ones of an ideal fluid~\cite{Landau1987Fluid}}.
Moreover, the matching condition in Eqs.~(\ref{eq:matching1}) and~(\ref{eq:matching_f}) allow one to define an effective temperature {$T$ and quark-flavor chemical potentials $\mu_f$} for the out-of-equilibrium fluid. 
In the one-fluid approximation, the temperature $T$ is assumed to be common {to all fluid components (quarks, antiquarks and gluons in this case).
{The choice of the Landau-Lifshitz frame~\cite{Landau1987Fluid}
\begin{equation}
u_\nu T^{\mu \nu}_{\rm m} = - \varepsilon \hspace{0.03cm} u^{\mu}    
\end{equation}
corresponds to defining the collective fluid velocity as the eigenvector of the fluid energy-momentum tensor with the energy density $\varepsilon$ as its eigenvalue.

The conservation law for the net-quark current gets a correction due to the non-ideal term in Eq.~(\ref{eq:diff-f}):
\begin{equation}
    Dn_f+n_f\Theta+\nabla_\mu j_f^\mu=0\,.\label{eq:non-ideal-diffusion}
\end{equation}
Now we focus on the energy-momentum balance in the presence of the Lorentz force
acting on the fluid, possibly in a curved background. 
{Concerning energy conservation, by projecting Eq.~(\ref{eq:dmuTmunu-matt}) along the fluid four-velocity one gets:}
\begin{equation}
D \varepsilon + \left(\varepsilon + P + \Pi \right) \Theta + \pi^{\mu\nu} \sigma_{\mu \nu} = e^{\mu} \, (j_Q)_{\mu}  \label{eq:en}\,,
\end{equation}
{where the RHS represents the effect of the Joule heating, which contributes to slowing down the cooling of the fireball, together with the other hydrodynamic dissipative processes. Notice that, being $e^\mu$ orthogonal to the fluid four-velocity, only the dissipative contribution to the electric current appears in the RHS.
In the above the shear tensor of Eq.~(\ref{eq:shear}) has been introduced.
On the other hand, as in the ideal case, taking the projection of Eq.~(\ref{eq:dmuTmunu-matt}) orthogonal to the velocity one gets the equation for the fluid acceleration, namely:}
\begin{multline}
 (\varepsilon +P+\Pi) \, a_\mu + \Delta_\mu (P+\Pi) + \Delta_{\mu\beta} \, \Delta_\alpha \pi^{\alpha\beta} + 
a^\nu\,\pi_{\nu\mu}=\\
= n_Q \, e_\mu + \epsilon_{\mu\nu\lambda\rho} \, j_Q^\nu \, b^\lambda \, u^\rho \label{eq:mom} \,,
\end{multline}
{whose RHS contains the effect of the Lorentz force.}

In the resistive case, the energy-momentum tensor of the EM fields {is given by}~\cite{Misner:1973prb,Anile1990}:
\begin{align}
\nonumber
    T^{\mu\nu}_{\rm em} = & \, \frac{1}{2} \left( e^2 + b^2 \right) u^\mu u^\nu + \frac{1}{2} \left( e^2 + b^2 \right) \Delta^{\mu\nu} - \\ 
    & - e^\mu e^\nu - b^\mu b^\nu + 2 \, S^{(\mu} u^{\nu)} \,.
\end{align}
In the RHS of the above relation, $S^\mu \equiv \epsilon^{\mu\alpha\beta\gamma} e_\alpha \, b_\beta \, u_\gamma$ has the interpretation of EM flux of energy and reduces to the well-known Poynting's three-vector in the fluid LRF.

{Finally, dissipative processes introduce a correction to the entropy current,}
\begin{equation}
\mathcal{S}^{\,\mu}  =  s u^\mu + \delta \mathcal{S}^{\,\mu} \,, \label{eq:entropy_nonid}
\end{equation}
whose four-divergence -- exploiting Eqs.~(\ref{eq:Euler-multiple}) and~(\ref{eq:1stlaw-multiple}) -- can be written as
\begin{equation}
\nabla_\mu \mathcal{S}^\mu = \frac{1}{T}D\varepsilon + \frac{\varepsilon+p}{T}\Theta + \sum_{f}\frac{\mu_f}{T}\nabla_\mu j_f^\mu +\nabla_\mu \, \delta\mathcal{S}^\mu \,.
\end{equation}

One is left with the problem of finding constitutive relations to fix the dissipative terms $\Pi$, $\pi^{\mu\nu}$ and $j_f^\mu$ entering the conservation laws in Eqs.~(\ref{eq:non-ideal-diffusion}),~(\ref{eq:en}) and~(\ref{eq:mom}). In this section we follow the macroscopic approach of imposing through Eq.~(\ref{eq:2ndlaw}) a non-negative rate of entropy production. At first order in a gradient expansion one has
\begin{equation}
\delta \mathcal{S}^\mu \equiv - \sum_f^{u, d, s}\alpha_f \, j_f^\mu \,, \label{eq:s1order-multiple}    
\end{equation}
leading to
\begin{multline}
T \nabla_\mu \mathcal{S}^\mu = 
- \Pi \,\Theta - \pi^{\mu\nu}\sigma_{\mu\nu}+
\sum_f^{u, d, s}\left(\mu_f - \alpha_f T \right) \nabla_\mu j_f^\mu+\\
+\sum_f^{u, d, s} j_f^\mu\left[\varmathbb{Q}_f \, |e| \, e_\mu-T\Delta_\mu\left(\frac{\mu_f}{T}\right)\right]\ge 0\,.
\label{eq:Srate-3f}
\end{multline}
A sufficient condition for the latter to be satisfied is that $\alpha_f=\mu_f/T$, the viscous corrections are given by
    \begin{equation}
      \Pi\! \equiv\! -\zeta\Theta\,,\qquad
      \pi^{\mu\nu} \!\equiv\! -2\eta \,\sigma^{\mu\nu}\,,
\end{equation}
and
\begin{equation}
{j_f^\mu \equiv \kappa'_{ff'} \left(-T \Delta^\mu \alpha_{f'}+\varmathbb{Q}_{f'} \, |e| \, e^\mu\right)}\label{eq:jdiff_ff'}
\end{equation}
so that
    \begin{equation}
T \nabla_\mu \mathcal{S}^\mu =  \frac{\Pi^2}{\zeta} + \frac{\pi^{\mu\nu}\pi_{\mu\nu}}{2\eta}
+ j_f^\mu\,\bigl(\kappa'^{-1}\bigr)_{ff'}\,(j_{f'})_\mu\ge 0\,,\label{eq:entropy-production}     
      \end{equation}
with positive bulk and shear viscosity $\zeta,\eta >0$ and the quark flavor-diffusion matrix $\kappa'_{ff'} \equiv \kappa_{ff'}/T$ being positive-definite. Notice that the dissipative flavor-current in Eq.~(\ref{eq:jdiff_ff'}) contains both a diffusion and a conductive contribution, with the same transport coefficients describing the response both to a density gradient and to an electric field. It is convenient to express it in the macroscopic-charge basis ($q={\cal B},{Q},S$):
    \begin{equation}
      j^\mu_{q}=-T  \underbrace{{\cal M}_{qf}\,\kappa'_{ff'}\,({\cal M}^{\rm T})_{f'q'}}_{\displaystyle{\equiv\kappa'_{qq'}}}\Delta^\mu \alpha_{q'}+
      \underbrace{{\cal M}_{qf}\,\kappa'_{ff'}\,\varmathbb{Q}_{f'}\,|e|}_{\displaystyle{\equiv\sigma_q}}\, e^\mu\,.\label{eq:WF0}
    \end{equation}
Setting $q=Q$ one gets a generalized {Wiedemann-Franz law}    
    \begin{equation}
{\sigma_Q}=\sum_{f,f'}|e|^2\varmathbb{Q}_f\, \kappa'_{ff'}\,\varmathbb{Q}_{f'}\equiv \sum_{f,f'}|e|^2\frac{\varmathbb{Q}_f\, {\kappa_{ff'}}\,\varmathbb{Q}_{f'}}{T}\label{eq:WF}
\end{equation}
connecting the flavor-diffusion matrix with the electric conductivity. Notice that -- Eq.~(\ref{eq:WF0}) holding for all conserved charges-- in response to an electric field one gets also a non-vanishing baryon and strangeness conduction current.

Considering Eqs.~(\ref{eq:jdiff_ff'}) and~(\ref{eq:entropy-production}) it is worth stressing that, if a small entropy production has to be associated to a large (although finite) flavor-diffusion/charge-conductivity coefficients, it is necessary to have $e^\mu\sim \Delta^\mu\alpha_f$, with both quantities being small. Hence, in the following, the electric field in the LRF of the plasma will be considered a perturbative first-order quantity within a gradient expansion.  
\section{Microscopic derivation of the dissipative currents: the Boltzmann-Vlasov equation}
\label{sec:BV}
Here we are interested in providing a microscopic derivation of the conserved currents (the electric one, in particular) carried by quarks and antiquarks, in order to fix through a first-principle calculation the transport coefficients involved in the response of the fluid to local density gradients and/or electric fields. 
The starting point is the Boltzmann-Vlasov (BV) equation for the (anti-)particle distribution $\fpmf$, here written in the relaxation-time approximation and, for the sake of simplicity, for a flat spacetime in Minkowskian coordinates:
\begin{equation}
\left[p^\mu\partial_\mu+\Qpmf\,|e|\,F^{\mu\nu}\,p_\nu\frac{\partial}{\partial p^\mu}\right]f_f^\pm=\frac{p\cdot u}{\tau_R}(f^\pm_f-\fpmfo)\,.\label{eq:BVpm}   
\end{equation}
In the above $\Qpmf\!\equiv\!\pm\,\varmathbb{Q}_f$, $\fpmfo$ is a local thermal equilibrium distribution and $\tau_R$ is a macroscopic relaxation time, for simplicity taken as common to all particle species.
In the QGP only quarks and antiquarks are directly coupled to the electromagnetic field and are carriers of globally conserved charges. Hence in the following we will neglect gluons, even if their contribution to the energy-momentum tensor (including its dissipative corrections) and to the EoS is obviously taken into account.
One can recast Eq.~(\ref{eq:BVpm}) into the form
\begin{equation}
   \fpmf=\fpmfo+\frac{\tau_R}{p\!\cdot\! u}\left[p^\mu\partial_\mu+\Qpmf\,|e|\,F^{\mu\nu}\,p_\nu\frac{\partial}{\partial p^\mu}\right]f_f^\pm\,,\label{eq:iterative} 
\end{equation}
allowing to formally express its solution as a geometric series
\begin{equation}
\fpmf=\sum_{n=0}^\infty\left[\frac{\tau_R}{p\cdot u} \left(p^\mu\partial_\mu+\Qpmf\,|e|\,F^{\mu\nu}\,p_\nu\frac{\partial}{\partial p^\mu} \right)\right]^n\fpmfo\,.\label{eq:BV}
\end{equation}
Notice that, while Eq.~(\ref{eq:BVpm}) is very general and applicable to situations in which the plasma is far from equilibrium, both the evaluation of transport coefficients and the derivation of relativistic MHD equations are based on the truncation of the above expansion at some finite order. Three independent conditions have to be fulfilled in order such a truncation to be meaningful.
\begin{enumerate}
    \item ${\rm Kn}\equiv\lambda_{\rm mfp}/L\sim|\tau_R\,\partial|\ll 1$: in the absence of electromagnetic fields this would correspond to the usual Chapman-Enskog gradient expansion, valid for small values of the Knudsen number Kn.
    \item $\xi\equiv\tau_R\,|e|\,E/T\ll 1$: the energy gained from the electric field by the charged plasma particles between two collisions has to be small compared to the thermal energy.
    \item $\chi\equiv\tau_R\,|e|\,B/T\sim\lambda_{\rm mfp}/r_{\rm Larm}\ll 1$: the mean free path of the charged plasma particles has to be small compared to their Larmor radius in the magnetic field; the bending of the trajectories between two collisions has to be negligible.  
\end{enumerate}
In the following we work under the assumption that the first and second condition are satisfied: in particular, as expected for a good electric conductor (see the discussion at the end of Sec.~\ref{sec:dissipative}), the local electric field is treated as a perturbative quantity of first order in the gradients. On the other hand, we let the magnetic field vary over a wide range, so that the third condition may be not necessarily satisfied. In this case a self-consistent treatment of magnetic effects is necessary.

We aim to connect the parameter $\chi$ to other standard dimensionless coefficients used in plasma physics, where the magnetic vs matter dominance is usually quantified, depending on the case, by the so called inverse beta-value $\beta_V^{-1}$ and magnetization $\sigma$, defined as:
\begin{equation}
    \beta_V^{-1}\equiv\frac{b^2/2}{P}\quad{\rm and}\quad \sigma\equiv\frac{b^2}{\varepsilon+P}\,.\label{eq:beta-inv}
\end{equation}
The inverse beta-value represents the ratio between the magnetic and thermal pressure in LRF of the fluid. In cold astrophysical plasmas the thermal pressure may be negligible and the energy density may be dominated by the mass density. In this case the magnetic pressure is more conveniently normalized to the enthalpy density. Strongly-magnetized plasmas are very common in astrophysics, for instance in jets following a binary neutron-star merger event~\cite{Mattia:2024erd}, in pulsar magnetospheres~\cite{Belyaev:2014fla,2013MNRAS.435L...1T} and in the solar atmosphere~\cite{Bourdin_2017}. Concerning the environment produced in relativistic HIC's a back-of-the-envelope estimate can be attempted. From\footnote{Here we assumed that the relaxation time receives contribution mainly from strong interactions among partons, dominant over electromagnetic interactions. This certainly holds true in the weak-field limit addressed in Sec.~\ref{sec:weak}. However, in a completely rigorous treatment of the strongly-magnetized regime, $\tau_R$ can get modified such that the ratio $\eta_{\perp}/s$ never becomes smaller than the KSS lower bound provided by holographic calculations~\cite{Dey:2019axu,Ammon:2020rvg,Finazzo:2016mhm}.}
\begin{equation}
\tau_R=5\,(\eta/s)\frac{1}{T}\,,\qquad P=\frac{g_{\rm dof}}{\pi^2}T^4\,,\qquad\beta_V\equiv \frac{P}{b^2/2}\,,\label{eq:BOE}
\end{equation}
assuming {$\eta/s\approx 0.2$} for the specific viscosity and {$g_{\rm dof}\approx 50$} for the effective number of relativistic degrees of freedom, one gets
    \begin{equation}
    \label{eq:beta_chi}
{\chi^2}=\frac{50 \,g_{\rm dof}\,(\eta/s)^2\,4\pi\,\alpha_{\rm em}}{\pi^2} \, \beta_V^{-1}\,{\approx \beta_V^{-1}}\,.      
    \end{equation}
Considering for instance Fig.~10 of Ref.~\cite{Inghirami:2016iru}, referring to a peripheral Au-Au collision at $\sqrt{s_{\rm NN}}=200$ GeV, one can see that the bulk of the produced fireball at the beginning of its evolution is characterized by $\beta_V^{-1}<10^{-2}$, ensuring $\chi\ll 1$. On the other hand, in the most peripheral regions, the inverse beta-value can become of order 1 or even larger. Hence, providing a macroscopic derivation of flavor/charge transport coefficients in a strongly-magnetized plasma can be relevant for a realistic modeling of the fireball and magnetic-field evolution.

The starting point of the calculation is the expression of the dissipative quark-flavor current in terms of the off-equilibrium fluctuations of the quark and antiquarks one-particle distributions:
\be
j_f^\mu\equiv g_f\,\Delta^\mu_{\,\,\,\,\nu} \int d\chi\,p^\nu\,\left[\delta f_f^+-\delta f_f^-\right]\,,\quad{\rm with}\quad d\chi\!\equiv\!\frac{d^3 p}{(2\pi)^3}\frac{1}{\epsilon_p}\,,\label{eq:jdiss-def}
\ee
where, within the RTA approximation, 
\be
\dfpmf=\frac{\tau_R}{p\!\cdot\! u}\left(p^\mu\partial_\mu+\Qpmf\,|e|\,F^{\mu\nu}\,p_\nu\frac{\partial}{\partial p^\mu}\right)\fpmf\,.\label{eq:deltaf-full}    
\ee
We consider the ideal EoS of a conformal plasma of massless gluons and 3 light quark flavors ($f=u,d,s$), whose thermal pressure is given by~\cite{Jaiswal:2015mxa}
    \begin{equation}
P=\beta^{-4}\left[\frac{(4g_g+7 \sum_f g_f)\pi^2}{360}+\sum_f\left(\frac{g_f}{12}\,\alpha_f^2+\frac{g_f}{24\pi^2}\,\alpha_f^4\right)\right]\,.\label{eq:P-conformal}      
    \end{equation}
In the above $g_g\!=\!2\times 8$ accounts for the polarization and colors of gluons, while $g_f\!=\!2\times 3$ (equal for all flavors) refers to the spin and colors of quarks.
We now consider the conservation laws for quark flavors energy and linear momentum given in Eqs.~(\ref{eq:non-ideal-diffusion}),~(\ref{eq:en}) and~(\ref{eq:mom}). In the case of a conformal plasma, in which the temperature is the only physical scale, and neglecting subleading corrections in Kn and $\xi$ they reduce to the following much simpler form:
{\setlength\arraycolsep{1pt}    
    \begin{eqnarray}
      D\,\alpha_f&=&0\nonumber\\
      D\,\beta&=&\displaystyle{\frac{1}{3}}\,\beta\,\Theta\nonumber\\
      \Delta^\mu\beta&=&\beta\, a^\mu+\sum_f\frac{n_f}{\varepsilon+P}\,(\Delta^\mu\alpha_f-\beta\,\varmathbb{Q}_f\,|e|\,e^\mu)\,-\nonumber\\
      {}&{}&-\frac{\beta}{\varepsilon+P}\sum_f\varmathbb{Q}_f\,|e|\,\epsilon^{\mu\nu\lambda\rho}\,(j_f)_\nu \,b_\lambda \,u_\rho\,.\label{eq:set-conformal}
      \end{eqnarray}}
\subsection{Weakly-magnetized QGP} \label{sec:weak}
 In a weakly-magnetized plasma, beside having ${\rm Kn},\,\xi\ll 1$, the condition $\chi\ll 1$ is also fulfilled. Hence in the RHS of Eq.~(\ref{eq:deltaf-full}) one can substitute $\fpmf\to\fpmfo\equiv f_0(\xpmf)$, where
 \begin{equation}
     \fpmfo
     ={\frac{1}{\displaystyle{e^{\xpmf}+1}}}\,,\quad
     \xpmf\equiv -\beta\,(p\cdot u)-\apmf\quad{\rm and}\quad \apmf\equiv\pm\alpha_f\,,
 \end{equation}
 since off-equilibrium corrections would give rise to terms at least of second order in our small expansion parameters. Defining $\widetilde\fpmfo\equiv1-\fpmfo $ and taking into account that 
 \begin{align}
     {p^\mu\partial_\mu \fpmfo}&=\frac{d f_0}{d\xpmf}\,{p^\mu\partial_\mu \xpmf}=\left[-\fpmfo\widetilde\fpmfo\right]\,{p^\mu\partial_\mu \xpmf}\nonumber\\
\frac{\partial \fpmfo}{\partial p^\mu}&=
\frac{d f_0}{d \xpmf}\,\frac{\partial\xpmf}{\partial p^\mu}
=\left[-\fpmfo\widetilde\fpmfo\right](-\beta \,u_\mu) 
 \end{align}
one gets 
  {\setlength\arraycolsep{1pt}    
    \begin{eqnarray}  
 \dfpmf&=&\frac{\tau_R}{p\!\cdot\! u}\left(p^\mu\partial_\mu+\Qpmf\,|e|\,F^{\mu\nu}\,p_\nu\frac{\partial}{\partial p^\mu}\right)\fpmfo\nonumber\\
{}&=&\frac{\tau_R}{p\!\cdot\! u}\left[-\fpmfo\,\widetilde\fpmfo\right]\Bigl(\underbrace{p^\mu\partial_\mu \xpmf}_{I} \underbrace{+\,\beta\,\Qpmf\,|e|\,p^\nu\,e_\nu}_{II} \Bigr)\,,
  \end{eqnarray}}
where the magnetic field does not contribute, due to the isotropy of the local equilibrium distributions. We now focus on the evaluation of contribution $I$. Exploiting the conservation laws for a conformal system in Eq.~(\ref{eq:set-conformal}) one gets:
\begin{multline}
p^\mu\partial_\mu \xpmf= \left[(p\!\cdot\! u)\,D-p^\rho\Delta_\rho\right]\,[\beta\,(p\!\cdot\! u)+\apmf]=\\
=\underbrace{-\,p^\rho\Delta_\rho\apmf}_{Ia}\underbrace{-\,(p\!\cdot\! u)\,p^\rho \sum_{f'}\frac{n_{f}'}{\varepsilon+P}\,\left(\Delta_\rho\alpha_{f'}-\beta\,\varmathbb{Q}_\ff\,|e|\,e_\rho\right)}_{Ib}+\,...\, 
\end{multline}
where we here and in the following we keep only the terms providing a non-zero contribution to the dissipative flavor current in Eq.~(\ref{eq:jdiss-def}). Putting together the terms $II$, $Ia$ and $Ib$ the off-equilibrium fluctuation of the one-particle distribution in a weakly-magnetized plasma reads then
{\setlength\arraycolsep{1pt}
    \begin{eqnarray}
\dfpmf&=& p^\rho{\frac{\tau_R}{(-p\!\cdot\! u)}\left[\fpmfo\widetilde\fpmfo\right]\left(\pm\delta_{ff'}-\frac{n_{f'}\,(-p\!\cdot\! u)}{\varepsilon+P}\right)}\,
g_{\rho\sigma}\times\nonumber\\
{}&{}&\times
\,{\left(\,\beta\,\varmathbb{Q}_{f'}|e|\,e^\sigma-\Delta^\sigma\alpha_{f'}\right)}= p^\rho\,A^\pm_{ff'}\,\Delta_{\rho\sigma}\,\widetilde e^{\,\sigma}_\ff\label{eq:df-weak}
    \end{eqnarray}}
where we defined
\begin{equation}
A^\pm_{ff'}\equiv \frac{\tau_R}{(-p\!\cdot\! u)}\left[\fpmfo\widetilde\fpmfo\right]\left(\pm\delta_{ff'}-\frac{n_{f'}\,(-p\!\cdot\! u)}{\varepsilon+P}\right)\,,\label{eq:A_ff'}    
\end{equation}
\begin{equation}
\widetilde e^{\,\sigma}_\ff\equiv \left(\,\beta\,\varmathbb{Q}_{f'}|e|\,e^\sigma-\Delta^\sigma\alpha_{f'}\right)\,\label{eq:etilde-def}    
\end{equation}
and we exploited $u_\mu e^\mu=u_\mu\Delta^\mu=0$ to replace $g_{\rho\sigma}\to\Delta_{\rho\sigma}$.
In the above $(-p\!\cdot\! u)\!\equiv\!\epsilon_p^*$ represents the quark energy in the plasma LRF.
Notice, first of all, that Eqs.~(\ref{eq:df-weak}) and~(\ref{eq:A_ff'}) entail that, in the LRF of a weakly-magnetized plasma, the response of the particle distribution to an electric field or a density gradient is isotropic, independent of the orientation of the magnetic field. For the induced flavor current one gets:
{\setlength\arraycolsep{1pt}
\begin{eqnarray}
    j^\mu_f&=&g_f\,\Delta^\mu_{\,\,\,\,\nu} \int d\chi\,\,p^\nu p^\rho\,\left(A^+_{ff'}-A^-_{ff'}\right)\Delta_{\rho\sigma}\,\widetilde e^{\,\sigma}_\ff\nonumber\\
    {}&=&g_f\,\frac{1}{3}\int d\chi\,\left(\Delta_{\nu\rho}\,p^\nu p^\rho\right)\,\left(A^+_{ff'}-A^-_{ff'}\right)\,\Delta^\mu_{\,\,\sigma}\,\widetilde e^{\,\sigma}_\ff\equiv \kappa_{ff'}\,\widetilde e^{\,\mu}_\ff\nonumber\\
\end{eqnarray}
}
Also for the induced current one gets an isotropic response in the limit of weak magnetization. The flavor-diffusion matrix can be readily identified:
{\setlength\arraycolsep{1pt}
\begin{eqnarray}
\kappa_{ff'}&=&
g_f\,\frac{1}{3}\int d\chi\,\left(\Delta_{\nu\rho}\,p^\nu p^\rho\right)\,\left(A^+_{ff'}-A^-_{ff'}\right)\nonumber\\
{}&\underset{\rm LRF}{=}&g_f\,\frac{1}{3}\int d\chi\,\, \vec p^{\,2}  \left(A^+_{f\ff}\!-\!A^-_{f\ff} \right)\,,\label{eq:kappa-weak1}
\end{eqnarray}}
where the first line is valid in an arbitrary frame, while the last one refers to the LRF in which $\Delta_{\nu\rho}p^\nu p^\rho=\vec p^{\,2}$. Details on the evaluation of the integrals in Eq.~(\ref{eq:kappa-weak1}) are provided in~\ref{App:integrals}. Eventually one obtains
\begin{equation}
    \kappa_{ff'}=\tau_R\left[\frac{g_f T^3}{18}\left(1+\frac{3}{\pi^2}\alpha_{f}^2\right)\delta_{ff'}-\frac{n_f\, n_{f'}T}{\varepsilon+P}\right]\,.\label{eq:kappa-weak}
\end{equation}
Hence, at finite density and in the weak field limit, flavor diffusion is described by a symmetric matrix with non-vanishing off-diagonal elements introducing a coupling between different quark species, as expected from Onsager's relations~\cite{Fotakis:2019nbq,Gavassino:2023qnw}. On the other hand, for zero net-quark density the flavor-diffusion matrix is purely diagonal.

It is interesting to compare the above results with current lattice-QCD estimates, so far limited to the case of vanishing quark chemical potentials.
Setting $g_f\!=\!6$ in Eq.~(\ref{eq:kappa-weak}) and  exploiting the Wiedemann-Franz law in Eq.~(\ref{eq:WF}) kinetic theory predicts
\begin{equation}
      \sigma_Q=\sum_{f,f'}|e|^2\frac{\varmathbb{Q}_f\,{\kappa_{ff'}}\,\varmathbb{Q}_{f'}}{T}=\frac{\tau_R}{T}\frac{T^3}{3}\,{|e|^2\sum_f\varmathbb{Q}_f^2}\equiv {\tau_R}\frac{T^2}{3}\,{C_{\rm em}}    
\end{equation}
for the electric conductivity in the limit of vanishing net-quark density. In the above the trivial dependence on the quark electric charge has been absorbed in the factor $C_{\rm em}\equiv |e|^2\sum_f\mathbb{Q}_f^2$. This makes it possible to perform a meaningful comparison with lattice-QCD simulations referring to different numbers of light quark flavors. Expressing the relaxation time in terms of the specific viscosity as in Eq.~(\ref{eq:BOE}) one gets:
\begin{equation}
\frac{\sigma_Q}{T\,C_{\rm em}}=\frac{5}{3}\left(\eta/s\right)\,.
\end{equation}
For realistic values of the specific viscosity $0.1\,\lsim\,\eta/s\,\lsim\, 0.2$ this results is in remarkable quantitative agreement with current lattice-QCD estimates. Also the growing behavior of $\sigma_Q/T$ with the temperature found on the lattice looks consistent~\cite{Aarts:2020dda,Aarts:2014nba,Brandt:2015aqk,Astrakhantsev:2019zkr} with the results for the QGP specific shear-viscosity coming from Bayesian analysis~\cite{Moreland:2018jos,Parkkila:2021tqq}. 
\subsection{Strongly-magnetized QGP}
\label{sec:strong}
When the magnetic parameter $\chi$ becomes of order 1, truncating the expansion in Eq.~(\ref{eq:BV}) is no longer possible. In fact, all terms of the form
\begin{multline*}
\left[\left(\frac{\tau_R}{p\!\cdot\! u}\right)\left(\Qpmf\,|e|\,b^{\mu\nu}\,p_\nu\frac{\partial}{\partial p^\mu}\right)\right]^n
\left[ \left(\frac{\tau_R}{p\!\cdot\! u}\right)\left(p^\mu\partial_\mu+\right.\right.\\
\left.\left.+\Qpmf\,|e|\,F^{\mu\nu}\,p_\nu\frac{\partial}{\partial p^\mu}\right)\right]\fpmfo
\end{multline*}
would provide a contribution of first order in the small expansion parameters ${\rm Kn}$ and $\xi$. It is then more convenient to look for an approximate self-consistent solution of Eq.~(\ref{eq:BVpm}), in which space-time gradients and the electric field are taken as perturbative quantities, while the magnetic field can be large.
The equation consistently including all first-order terms in Kn and $\xi$ is
\begin{multline}
\frac{\tau_R}{p\!\cdot\! u}\left[p^\mu\partial_\mu+ \Qpmf\,|e|\, (e^\nu p_\nu)\, u^\mu\,\frac{\partial}{\partial p^\mu} \right]\fpmfo+\\
+\frac{\tau_R}{p\!\cdot\! u}\,\Qpmf\,|e|\, b^{\mu\nu}\,p_\nu \,\frac{\partial}{\partial p^\mu}\dfpmf=\dfpmf\,,\label{eq:selfcons}
\end{multline}
where the absence of the magnetic field in the first line is due to the isotropy of the equilibrium one-particle distribution in the LRF.  Inspired by Refs.~\cite{Harutyunyan:2016rxm,Feng:2017tsh} we now make the following ansatz on the form of $\dfpmf$, keeping only the terms providing a non-vanishing contribution to the dissipative net-quark current:
\begin{equation}
\label{eq:ansatz}
\dfpmf=p^\rho\left[E^\pm_{ff'}\,\Delta_{\rho\sigma}+ B^\pm_{ff'}\,\hat b_\rho \hat b_\sigma+ H^\pm_{ff'}\,\hat b_{\rho\sigma}\right]\,\widetilde e^{\,\sigma}_\ff\,.
\end{equation}
In the above equation $\hat b_\rho\!\equiv\! b_\rho/B$ and $\hat b_{\rho\sigma}\!\equiv\! \epsilon_{\rho\sigma\lambda\kappa}\hat b^\lambda u^\kappa$, with $B\!\equiv\!\sqrt{b^2}$. One can conveniently introduce the following projector
\begin{equation}
    \Xi^{\mu\nu}\equiv\Delta^{\mu\nu}-\hat b^\mu\hat b^\nu\,,
\end{equation}
which is orthogonal both to the fluid velocity $u^\mu$ and to the magnetic field, as can be easily checked. In the LRF, its only non-vanishing components read
\begin{equation}
     \Xi^{ij}\equiv\delta^{ij}-\hat b^i \hat b^j\,.
\end{equation}
and the latter represents the projection operator on the plane orthogonal to the magnetic field.
Furthermore, the following identities hold:
\begin{equation}
    \Xi^{\mu\nu}\Xi_{\mu\nu}=2\,,\quad
    \hat b^{\mu\nu}\hat b_{\mu\nu}=2\,,\quad
    \hat b^{\mu\alpha}\hat b_{\nu\alpha}=\Xi^{\mu}_{\,\,\,\,\nu} \,.
\end{equation}
Accordingly, the fluctuation of the single-particle distribution can equivalently be rewritten as
\begin{equation}
\dfpmf=p^\rho\left[E^\pm_{ff'}\,\Xi_{\rho\sigma}+L^\pm_{ff'}\,\hat b_\rho \hat b_\sigma+ H^\pm_{ff'}\hat b_{\rho\sigma}\right]\,\widetilde e^{\,\sigma}_\ff\,,\label{eq:dfcovproj}
\end{equation}
where we introduced the notation $L^\pm_{ff'}\equiv B^\pm_{ff'}+E^\pm_{ff'}$ for the longitudinal response to an electric-field (or density gradient). In the above $E^\pm_{ff'}$ and $H^\pm_{ff'}$ represent instead the transverse and Hall responses.
Our goal now is to explicitly determine $E^\pm_{ff'}$, $L^\pm_{ff'}$ and $H^\pm_{ff'}$ by substituting the above expression for $\dfpmf$ into Eq.~(\ref{eq:selfcons}).
Without loss of generality, the calculation can be performed in the local rest frame of the fluid, where $e^0\!=\!b^0\!=\!0$ and $\hat b_{\rho\sigma}\!=\!\epsilon_{\rho\sigma\lambda 0}\,\hat b^\lambda$ and, accordingly,
\begin{equation}
    \left.\dfpmf\right|_{\rm LRF}=p^i\left[E^\pm_{ff'}\,\Xi^{ij}+ L^\pm_{ff'}\,\hat b^i \hat b^j+ H^\pm_{ff'}\,\epsilon^{ijk}\hat b^k\right]\,\widetilde e^{\,j}_\ff\,.
\label{eq:dfpmLRF}
\end{equation}
One gets then for the induced flavor current
\begin{multline}
j_f^{\,i}=g_f\!\int\!\!\frac{d^3 p}{(2\pi)^3}\frac{p^i}{\epsilon_p}\,p^l\,\left[(E^+_{f\ff}\!-\!E^-_{f\ff})\,\Xi^{lj}+\left(L^+_{f\ff}\!-\!L^-_{f\ff} \right)\hat b^l \hat b^j+\right.\\
+\left.(H^+_{f\ff}-H^-_{f\ff})\,\epsilon^{ljk}\hat b^k \right]\,\widetilde e^j_\ff\,\equiv\kappa_{ff'}^{ij}\,\widetilde e^j_\ff\,.\label{eq:jdiff-dec1}   
\end{multline}
The flavor-diffusion matrix acquires then a tensor structure given by:
\begin{equation}
\kappa_{ff'}^{ij}=\kappa^\perp_{ff'}\,\Xi^{ij}+\kappa^{||}_{ff'}\,\hat b^i\hat b^j+\kappa^\times_{ff'}\,\epsilon^{ijk}\hat b^k\,.\label{eq:jdiff-dec2}    
\end{equation}
Eq.~(\ref{eq:jdiff-dec2}) can be easily inverted obtaining
\begin{equation}
\kappa^{||}_{ff'}=\hat b^i \hat b^j \,\kappa_{ff'}^{ij}\,,\quad \kappa^{\perp}_{ff'}=\frac{1}{2}\,\Xi^{ij}\,\kappa_{ff'}^{ij}\,,\quad
\kappa^{\times}_{ff'}=\frac{1}{2}\,\epsilon^{ijm}\hat b^m\, \kappa_{ff'}^{ij}\,,
\end{equation}
leading to
\begin{subequations}
    \begin{align}
\kappa^{||}_{ff'}&=g_f\,\frac{1}{3}\int d\chi\, \vec p^{\,2}\,  \left(L^+_{f\ff}\!-\!L^-_{f\ff} \right)\,,\label{eq:kappa-L}\\
\kappa^{\perp}_{ff'}&=g_f\,\frac{1}{3}\int d\chi\, \vec p^{\, 2}\,  \left(E^+_{f\ff}\!-\!E^-_{f\ff} \right)\,,\label{eq:kappa-T}\\
\kappa^{\times}_{ff'}&=g_f\,\frac{1}{3}\int d\chi\, \vec p^{\, 2}\,  \left(H^+_{f\ff}\!-\!H^-_{f\ff} \right)\,,\label{eq:kappa-H}
    \end{align}
\end{subequations}
where the angular averages around the $\hat b$ direction
\begin{equation}
    \langle \cos^2\theta\rangle=\frac{1}{3}\quad{\rm and}\quad
    \langle 1-\cos^2\theta\rangle=\frac{2}{3}
\label{eq:angave}    
\end{equation}
have been exploited.
In the first line of Eq.~(\ref{eq:selfcons}) one can substitute the first-order result provided by Eqs.~(\ref{eq:df-weak}), however supplemented by an additional term arising from the spacetime gradient of the argument of $\fpmfo$
\begin{multline}
p^\mu\partial_\mu \xpmf=(Ia)+(Ib)+\\
+(p\cdot u)\,p_\mu\,\frac{\beta}{\varepsilon+P}\sum_\ff\varmathbb{Q}_\ff\,|e|\,\epsilon^{\mu\nu\lambda\rho}\,(j_\ff)_\nu \,b_\lambda u_\rho\,, 
\end{multline}
no longer sub-leading if the plasma is strongly magnetized. Notice that this last term arises from the contribution of the Lorentz force to the fluid acceleration appearing in Eqs.~(\ref{eq:mom}) and~(\ref{eq:set-conformal}). The latter was neglected in previous attempts of evaluating the charge diffusion and conductivity of a magnetized plasma~\cite{Harutyunyan:2016rxm,Feng:2017tsh,Dash:2020vxk,Dey:2019axu}. We will come back to this issue in discussing our numerical results. One gets:
\begin{multline}
\frac{\tau_R}{p\!\cdot\! u}\left[p^\mu\partial_\mu+ \Qpmf\,|e|\, (e^\nu p_\nu)\, u^\mu\,\frac{\partial}{\partial p^\mu} \right]\fpmfo=p^i\, A^\pm_{ff'}\,\delta^{ij}\,\,\widetilde e^j_\ff+\\
+\frac{\tau_R}{p\!\cdot\! u}\,\left[-\fpmfo\widetilde\fpmfo\right]\frac{\beta\,
(p\!\cdot\! u)}{\varepsilon+P} \,p^i\sum_{f''}\varmathbb{Q}_{f''}\,|e|\, B\,\epsilon^{ilk}j_{f''}^l\hat b^k \,,
\end{multline}
where in the RHS, written in the LRF, one has to substitute the previous decomposition for the induced current
\be
j^l_{f''}=\left[\kappa^\perp_{f''f'}\,\Xi^{lj}+\kappa^{||}_{f''f'}\,\hat b^l\hat b^j+\kappa^\times_{f''f'}\,\epsilon^{ljk}\hat b^k\right]\,\widetilde e^j_\ff\,,\label{eq:anstatz-strong}
\ee
leading to
\begin{multline}
\frac{\tau_R}{p\!\cdot\! u}\left[p^\mu\partial_\mu+ \Qpmf\,|e|\, (e^\nu p_\nu)\, u^\mu\,\frac{\partial}{\partial p^\mu} \right]\fpmfo=p^i\, A^\pm_{ff'}\,\delta^{ij}\,\,\widetilde e^j_\ff+\\
+p^i\,\sum_{f''}\frac{\beta\,\tau_R}{\varepsilon+P}\left[-\fpmfo\widetilde\fpmfo\right]\varmathbb{Q}_{f''}\,|e|\, B\,\times\\
\times\left[\kappa^{\perp}_{f''f'}\,\epsilon^{ijk}\hat b^k-\kappa^{\times}_{f''f'}\left(\delta^{ij}-\hat b^i \hat b^j\right)\right]\widetilde e^j_\ff\,.\label{eq:first-term}
\end{multline}
After introducing the matrix
\begin{equation}
    \overline{A}^{\,\pm}_{ff''}\equiv \frac{\beta\,\tau_R}{\varepsilon+P}\left[-\fpmfo\widetilde\fpmfo\right]\varmathbb{Q}_{f''}\,|e|\, B
\end{equation}
one can recast Eq.~(\ref{eq:first-term}) into the following more compact form:
\begin{multline}
\frac{\tau_R}{p\!\cdot\! u}\left[p^\mu\partial_\mu+ \Qpmf\,|e|\, (e^\nu p_\nu)\, u^\mu\,\frac{\partial}{\partial p^\mu} \right]\fpmfo=p^i\, A^\pm_{ff'}\,\Xi^{ij}\,\,\widetilde e^j_\ff+\\
+p^i\, A^\pm_{ff'}\,\hat b^i \hat b^j\,\widetilde e^j_\ff+
p^i\,\overline{A}^{\,\pm}_{ff''}\,\left[\kappa^{\perp}_{f''f'}\,\epsilon^{ijk}\hat b^k-\kappa^{\times}_{f''f'}\,\Xi^{ij}\right]\widetilde e^j_\ff\,.\label{eq:LHS-1st}
\end{multline}
We now address the second term in the LHS of Eq.~(\ref{eq:selfcons}), which for the comoving observer reads
\begin{multline}
-\frac{\tau_R}{\epsilon_p^*}\,\Qpmf\,|e|\,B\,\epsilon^{ilm}\hat b^m \,p^l\,\frac{\partial}{\partial p^i}\dfpmf= -\frac{\tau_R}{\tau_L} \,\Qpmf\,\epsilon^{ilm}\hat b^m \,p^l\times\\
\times\left[E^\pm_{ff'}\,\Xi^{ij}+ L^\pm_{ff'}\,\hat b^i \hat b^j+ H^\pm_{ff'}\,\epsilon^{ijk}\hat b^k\right]\,\widetilde e^j_\ff\,,
\end{multline}
where $\epsilon_p^*\equiv(-p\!\cdot\! u)$ is the single-particle energy in the LRF, $\tau_L\!\equiv\!\epsilon_p^*/\bigl(|e|\,B\bigr)$ represents the so-called Larmor timescale, which depends on the particle momentum, and $\dfpmf$ was decomposed as in Eq.~(\ref{eq:dfpmLRF}).
Performing the appropriate contractions one obtains:
\begin{multline}
-\frac{\tau_R}{\epsilon_p^*}\,\Qpmf\,|e|\,B\,\epsilon^{ilm}\hat b^m \,p^l\,\frac{\partial}{\partial p^i}\dfpmf=\\
=p^i\, \frac{\tau_R}{\tau_L} \,\Qpmf\,\left[ E^\pm_{ff'}\,\epsilon^{ijk}\hat b^k- H^\pm_{ff'}\,\Xi^{ij}\right]\,\widetilde e^j_\ff\,.\label{eq:LHS-2nd}
\end{multline}

After substituting Eqs.~(\ref{eq:LHS-1st}) and~(\ref{eq:LHS-2nd}) into Eq.~(\ref{eq:selfcons}) one finds the following results. First of all one has:
\begin{equation}
L^\pm_{ff'}=A^\pm_{ff'}\,.\label{eq:L}
\end{equation}
As expected from the nature of the Lorentz force, the longitudinal response to an electric field or a density gradient -- given by Eq.~(\ref{eq:kappa-L}) -- is not affected by the presence of a magnetic field.\\
In addition, the following coupled equations hold,
\begin{align*}
&A_{ff'}^\pm-\overline A^{\,\pm}_{ff''}\,\kappa^\times_{f''f'}-\frac{\tau_R}{\tau_L} \,\Qpmf\,H^\pm_{ff'}=E^\pm_{ff'}\,,\\
&\overline A^{\,\pm}_{ff''}\,\kappa^\perp_{f''f'}+\frac{\tau_R}{\tau_L} \,\Qpmf\,E^\pm_{ff'}=H^\pm_{ff'}\,,
\end{align*}
which allow one to determine the expressions for the ansatz parameters. Hence, one gets
\begin{subequations}
\begin{align}
  E^\pm_{ff'}&=\frac{\left(A_{ff'}^\pm-\overline A^{\,\pm}_{ff''}\kappa^\times_{f''f'}\right)-\Qpmf\,\frac{\tau_R}{\tau_L} \, \overline A^{\,\pm}_{ff''}\kappa^\perp_{f''f'}}{1+\left(\varmathbb{Q}_f \frac{\tau_R}{\tau_L}\right)^2}\label{eq:E}\\
H^\pm_{ff'}&=\frac{\Qpmf \,\frac{\tau_R}{\tau_L}\left(A_{ff'}^\pm- \overline A^{\,\pm}_{ff''}\kappa^\times_{f''f'}\right)+ \overline A^{\,\pm}_{ff''}\kappa^\perp_{f''f'}}{1+\left(\varmathbb{Q}_f \frac{\tau_R}{\tau_L}\right)^2}\,,\label{eq:H}
\end{align}
\label{eq:EHB}
\end{subequations}
The results in Eqs.~(\ref{eq:L}),~(\ref{eq:E}) and~(\ref{eq:H}) still depend on the components of the diffusion tensor and can be inserted into Eqs.~(\ref{eq:kappa-L}),~(\ref{eq:kappa-T}) and~(\ref{eq:kappa-H}).
Let us start from the longitudinal component. One gets
\begin{equation}
 \kappa_{ff'}^{||}=\frac{g_f}{3}\int\! d\chi\,{\vec p^{\,2}}\,\left(A^+_{f\ff}-A^-_{f\ff}\right)\,,\label{eq:kpar}    
\end{equation}
which coincides with the isotropic flavor-diffusion matrix in Eq.~(\ref{eq:kappa-weak1}) obtained in the limit of weak magnetization.\\
We now move to the evaluation of the transverse and Hall components of the flavor-diffusion matrix. We start isolating in Eqs.~(\ref{eq:E}) and~(\ref{eq:H}) the contributions not involving the $\overline{A}^{\,\pm}$ matrices. One has
\begin{subequations}
\begin{align}
\widetilde\kappa_{ff'}^{\,\perp}&=\frac{g_f}{3}\int\!d\chi\,{\vec p^{\,2}}\,\frac{A^+_{f\ff}-A^-_{f\ff}}{1+\left(\varmathbb{Q}_f \frac{\tau_R}{\tau_L}\right)^2}\,,\label{eq:kort}\\
\widetilde\kappa_{ff'}^{\,\times}&=\frac{g_f}{3}\int\! d\chi\,{\vec p^{\,2}}\,\frac{\left(\frac{\tau_R}{\tau_L}\right)\left(Q_f^+\,A^+_{f\ff}-Q_f^-\,A^-_{f\ff}\right)}{1+\left(\varmathbb{Q}_f \frac{\tau_R}{\tau_L}\right)^2}\label{eq:kx}\,, 
\end{align}
\end{subequations}
representing the net-quark diffusivities one would obtain by neglecting the contribution from the Lorentz force to the fluid acceleration, like in Refs.~\cite{Harutyunyan:2016rxm,Feng:2017tsh,Dash:2020vxk,Dey:2019axu}.

In the last integrand the following difference
\begin{multline}
Q_f^+ \,A^+_{f\ff}- Q_f^- \,A^-_{f\ff}= \varmathbb{Q}_f \frac{\tau_R}{\epsilon_p^*}\,\Bigg\{\left[\fpfo\widetilde\fpfo-\fmfo\widetilde\fmfo\right]\,\delta_{f\ff}\,-\Bigg.\\
\left.-\left[\fpfo\widetilde\fpfo+\fmfo\widetilde\fmfo\right]\,\frac{\epsilon_p^* \, n_\ff}{\varepsilon+P}
\right\}\,, \label{eq:kx-integrand}
\end{multline}
enters.
The full result instead reads:
\begin{multline}
\kappa_{ff'}^{\perp}=\widetilde\kappa_{ff'}^{\,\perp}-\frac{g_f}{3}\int\!d\chi\,{\vec p^{\,2}}\,\frac{\overline A^{\,+}_{ff''}-\overline A^{\,-}_{ff''}}{1+\left(\varmathbb{Q}_f \frac{\tau_R}{\tau_L}\right)^2}\,\kappa^\times_{f''f'}-\\
-\frac{g_f}{3}\int\! d\chi\,{\vec p^{\,2}}\,\frac{\left(\frac{\tau_R}{\tau_L}\right)\left(Q_f^+\,\overline A^{\,+}_{ff''}-Q_f^-\,\overline A^{\, -}_{f\ff}\right)}{1+\left(\varmathbb{Q}_f \frac{\tau_R}{\tau_L}\right)^2}\,\kappa^\perp_{f''f'}\\
\kappa_{ff'}^{\times}=\widetilde\kappa_{ff'}^{\,\times}-\frac{g_f}{3}\int\! d\chi\,{\vec p^{\,2}}\,\frac{\left(\frac{\tau_R}{\tau_L}\right)\left(Q_f^+\,\overline A^{\,+}_{ff''}-Q_f^-\,\overline A^{\, -}_{f\ff}\right)}{1+\left(\varmathbb{Q}_f \frac{\tau_R}{\tau_L}\right)^2}\,\kappa^\times_{f''f'}+\\
+\frac{g_f}{3}\int\!d\chi\,{\vec p^{\,2}}\,\frac{\overline A^{\,+}_{ff''}-\overline A^{\,-}_{ff''}}{1+\left(\varmathbb{Q}_f \frac{\tau_R}{\tau_L}\right)^2}\,\kappa^\perp_{f''f'}\,.
\end{multline}
After introducing the definitions
\begin{eqnarray}
\mathcal{T}_{ff''}&\equiv&\frac{g_f}{3}\int\!d\chi\,{\vec p^{\,2}}\,\frac{\overline A^{\,+}_{ff''}-\overline A^{\,-}_{ff''}}{1+\left(\varmathbb{Q}_f \frac{\tau_R}{\tau_L}\right)^2} \label{eq:Tdef}\,,\\
X_{ff''}&\equiv&\frac{g_f}{3}\int\! d\chi\,{\vec p^{\,2}}\,\frac{\left(\frac{\tau_R}{\tau_L}\right)\left(Q_f^+\,\overline A^{\,+}_{ff''}-Q_f^-\,\overline A^{\, -}_{f\ff}\right)}{1+\left(\varmathbb{Q}_f \frac{\tau_R}{\tau_L}\right)^2} \label{eq:Xdef} \,,
\end{eqnarray}
one eventually obtains that the following system of coupled equations
\begin{eqnarray}
\kappa_{ff'}^{\perp}&=&\widetilde\kappa_{ff'}^{\,\perp}-  \mathcal{T}_{ff''}\, \kappa_{f''f'}^{\times}-X_{ff''} \,\kappa_{f''f'}^{\perp}\,,\nonumber\\
\kappa_{ff'}^{\times}&=&\widetilde\kappa_{ff'}^{\,\times}-X_{ff''}\, \kappa_{f''f'}^{\times}+\mathcal{T}_{ff''} \,\kappa_{f''f'}^{\perp}\,,\label{eq:system}
\end{eqnarray}
must be satisfied, where there is an implicit sum over the flavor-index $f'' = u$, $d$, $s$. In the above one has
\begin{equation}
\overline A^{\,+}_{ff''}-\overline A^{\,-}_{ff''}=-\frac{\beta\,\tau_R}{\varepsilon+P}\left[\fpfo\widetilde\fpfo-\fmfo\widetilde\fmfo\right]\varmathbb{Q}_{f''}|e|\, B\label{eq:T-def} 
\end{equation}
and
\begin{equation}
Q_f^+\,\overline A^{\,+}_{ff''}-Q_f^-\,\overline A^{\,-}_{ff''}=-\frac{\beta\,\tau_R}{\varepsilon+P}\,\varmathbb{Q}_f\left[\fpfo\widetilde\fpfo+\fmfo\widetilde\fmfo\right]\varmathbb{Q}_{f''}|e|\, B\,. 
\end{equation}
As expected, in the absence of magnetic fields ($B\!=\! 0$ and, accordingly, $\tau_R/\tau_L\!\to\! 0$) one finds a vanishing Hall contribution, $\kappa_{ff'}^\times=\!\!0$, and equal diffusion coefficients along the transverse and longitudinal directions, $\kappa_{ff'}^\perp\! =\! \kappa_{ff'}^\parallel\!=\!\kappa_{ff'}$.
It is also interesting to study the behavior of the system in the limit of vanishing quark chemical potentials ($\alpha_f = 0$), but still in the presence of a large magnetic field, so that the present non-perturbative treatment is mandatory. First of all, notice that in the system of equations~(\ref{eq:system}), the mixing between the transverse and Hall components occurs through the matrix $\mathcal{T}_{ff''}$. However, the latter identically vanishes for zero quark chemical potential, as can be read from Eq.~(\ref{eq:T-def}). Hence, in this limit one has
\begin{eqnarray}
\kappa_{ff'}^{\perp}&=&[(\mathbf{1}+X)^{-1}]_{ff''}\, \widetilde\kappa_{f''f'}^{\,\perp}\nonumber\\  \kappa_{ff'}^{\times}&=&[(\mathbf{1}+X)^{-1}]_{ff''} \,\widetilde\kappa_{f''f'}^{\,\times}\,.\label{eq:zero-density} 
\end{eqnarray}
However, as one can see in Eq.~(\ref{eq:kx-integrand}), when all chemical potentials vanish one has $\widetilde\kappa_{f''f'}^{\,\times}\!=\!0$ and hence also $\kappa_{ff'}^{\,\times}\!=\!0$: no induced Hall current develops in the plasma. Nevertheless, if the latter is strongly magnetized, the transverse response is affected by the magnetic field and one has $\kappa_{ff'}^{\,\perp}\!\ne\! \kappa_{ff'}^{\,||}$.  

For the sake of simplicity and in order to more easily compare our findings with independent results obtained in the literature, we focus on the case of one single quark flavor. In this case, with obvious notation, the system in Eq.~(\ref{eq:system}) reduces to
\begin{eqnarray}
\kappa_{f}^{\perp}&=&\widetilde\kappa_{f}^{\,\perp}-  \mathcal{T}_{f}\, \kappa_{f}^{\times}-X_{f} \, \kappa_{f}^{\perp}\,,\nonumber\\
\kappa_{f}^{\times}&=&\widetilde\kappa_{f}^{\,\times}-X_{f} \,\kappa_{f}^{\times}+\mathcal{T}_{f} \,\kappa_{f}^{\perp}\,,\label{eq:system1f}
\end{eqnarray}
whose solutions are given by:
\begin{eqnarray}
\kappa_{f}^{\perp}&=& \frac{1}{\mathcal{T}^2_f + \left(1+X_f\right)^2} \, \Bigl[\left(1+X_f\right)\, \widetilde\kappa_{f}^{\,\perp} - \mathcal{T}_f \,\widetilde\kappa_{f}^{\,\times} \Bigr]\,,\label{eq:system1fs}\\
\kappa_{f}^{\times}&=&\frac{1}{\mathcal{T}^2_f + \left(1+X_f\right)^2} \, \Bigl[\left(1+X_f\right)\, \widetilde\kappa_{f}^{\,\times} + \mathcal{T}_f \,\widetilde\kappa_{f}^{\,\perp} \Bigr]\,.\label{eq:system1fs_re}
\end{eqnarray}
As in the case of multiple flavors, it is interesting to consider the limiting behavior of the above results for vanishing quark chemical potential, $\alpha_f=0$, but $B\ne 0$:
\begin{equation}   \kappa_{f}^{\perp}\,\Bigr|_{\alpha_f = 0}= \frac{\widetilde\kappa_{f}^{\,\perp}}{1+X_f}\Biggr|_{\alpha_f = 0} < \kappa_{f}^{\parallel}\quad,\quad \kappa_{f}^{\times}\,\Bigr|_{\alpha_f = 0}=0 \,.\label{eq:alpha0-1f}
\end{equation}
No Hall current develops in the absence of a net quark-number density; on the other hand, if $B$ is sufficiently large, the induced current in the plane transverse to the magnetic field is strongly reduced, at variance with what found, for instance, in Refs.~\cite{Panda:2021pvq,Panda:2020zhr}.  

\subsection{On the issue of entropy production}
In Sec.~\ref{sec:dissipative}, according to Eq.~(\ref{eq:Srate-3f}), the contribution to the entropy production rate from the induced quark-flavor currents was found to be proportional to $j^{\,\mu}_f\,\widetilde e_{\mu,f}$, with $\widetilde e_{\mu,f}$ defined in Eq.~(\ref{eq:etilde-def}). A sufficient condition to locally fulfill the second law of thermodynamics was to assume for the flavor current the form (the trivial spacetime tensor structure following from local isotropy in the weak-field limit and from the transverse nature of $\widetilde e^{\,\nu}_\ff$)
\begin{equation}  j^\mu_f=\kappa_{ff'}\,\Delta^\mu_{\,\,\,\nu}\,\widetilde e^{\,\nu}_\ff\,,\label{eq:ansatza-weak}
\end{equation}
with the flavor-diffusion matrix $\kappa_{ff'}$ having all positive eigenvalues, allowing one to get the non-negative result for the four-divergence of the entropy current in Eq.~(\ref{eq:entropy-production}).

However, in a strongly-magnetized plasma, the ansatz in Eq.~(\ref{eq:ansatza-weak}) is not general enough. In fact, the magnetic field breaks the isotropy of the system, so that -- as found in the previous section -- the most general expression for the induced flavor current (here written, without loss of generality, in the LRF) reads
\be
j^i_f=\kappa^{ij}_{ff'}\,\widetilde e^{\,j}_\ff\,,
\ee
with $\kappa^{ij}_{ff'}$ having the same tensor decomposition as in Eq.~(\ref{eq:jdiff-dec2}). For the entropy production rate this leads to
\be
\nabla_{\mu} \mathcal{S}^{\mu} = \ldots + j_f^{\,i}\,\bigl(\kappa^{-1}\bigr)_{f f'}^{il}\,j_{f'}^{\,l} \geq 0 \,.\label{eq:2ndlaw-strong}
\ee
Also the inverse diffusion tensor admits the decomposition
\begin{equation}
    \bigl(\kappa^{-1}\bigr)_{f f'}^{il} \equiv \mathcal{C}_{f f'}^{\perp}\, \Xi^{il} + \mathcal{C}_{f f'}^{\parallel} \,\Hat{b}^{\,i} \Hat{b}^{\, l} + \mathcal{C}_{f f'}^{\times} \, \Hat{b}^{il}\,,
\end{equation}
where we recall that $\Xi^{il} \equiv \delta^{il} - \Hat{b}^{\,i} \Hat{b}^{\,l}$ and $\Hat{b}^{il} \equiv \epsilon^{ilk}\,\hat{b}^{\,k}$. The next steps consist, first of all, in deriving the expression for $\mathcal{C}_{f f'}^{\perp}$, $\mathcal{C}_{f f'}^{\parallel}$ and $\mathcal{C}_{f f'}^{\times}$ by solving the equation
\begin{equation}
    \bigl(\kappa^{-1}\bigr)_{f f'}^{il} \, \kappa_{f' f''}^{lj} \equiv \delta_{f f''} \, \delta^{ij}\label{eq:inverse}
\end{equation}
and then in proving that the obtained results  satisfy Eq.~(\ref{eq:2ndlaw-strong}).

From Eq.~(\ref{eq:inverse}) one gets the following set of coupled algebraic equations
\begin{align}
\label{eq:system_inv1}
    & \mathcal{C}^{\perp}_{f f'} \, \kappa_{f' f''}^{\perp} - \mathcal{C}^{\times}_{f f'} \, \kappa^{\times}_{f' f''} = \delta_{f f''} \\
    \label{eq:system_inv2}
    & \mathcal{C}^{\parallel}_{f f'} \, \kappa^{\parallel}_{f' f''} - \mathcal{C}^{\perp}_{f f'} \, \kappa_{f' f''}^{\perp} + \mathcal{C}^{\times}_{f f'} \, \kappa^{\times}_{f' f''} = 0 \\
    \label{eq:system_inv3}
    & \mathcal{C}^{\perp}_{f f'} \, \kappa^{\times}_{f' f''} + \mathcal{C}^{\times}_{f f'} \, \kappa^{\perp}_{f' f''} = 0 \,.
\end{align}
By summing both sides of the first two equations one gets:
\begin{equation}
\label{eq:Cpar}
    \mathcal{C}^{\,\parallel}_{f f'} = \bigl(\kappa^{\,\parallel}\bigr)^{-1}_{f f'} \,.
\end{equation}
Notice that in the previous section -- see Eq.~(\ref{eq:kpar}) -- we found $\kappa^{\,\parallel}_{f\ff}=\kappa_{f\ff}$, with all positive eigenvalues, as required to satisfy Eq.~(\ref{eq:entropy-production}) in a weakly-magnetized plasma. This is also confirmed by an explicit numerical calculation (see Sec.~\ref{sec:num}).
For the other two components of the inverse diffusion tensor one gets
\begin{equation}
    \label{eq:Cperp}
    \mathcal{C}^{\perp}_{f f'} = \bigl(\kappa^{\perp}\bigr)^{-1}_{f f''} \, \Bigl[ \mathbf{1} + \Bigl(\kappa^{\times} \, \bigl(\kappa^{\perp}\bigr)^{-1} \Bigr)^2 \,\Bigr]^{-1}_{f'' f'}
\end{equation}
and
\begin{align}
    \label{eq:Ccross}
    \nonumber
    \mathcal{C}^{\times}_{f f'} &= - \,\mathcal{C}^{\perp}_{f f''} \, \kappa^{\times}_{f'' f'''} \, \bigl(\kappa^{\perp}\bigr)^{-1}_{f''' f'} = \\ 
    &= - \, \bigl(\kappa^{\perp}\bigr)^{-1}_{f \bar{f}} \, \Bigl[ \mathbf{1} + \Bigl(\kappa^{\times} \, \bigl(\kappa^{\perp}\bigr)^{-1} \Bigr)^2 \,\Bigr]^{-1}_{\bar{f} f''} \, \kappa^{\times}_{f'' f'''} \, \bigl(\kappa^{\perp}\bigr)^{-1}_{f''' f'}\,.
\end{align}
In order for the above expressions to be well defined the matrices in flavor space
\begin{equation*}
    \kappa^{\perp}\quad{\rm and}\quad
     \Bigl[\mathbf{1} + \Bigl(\kappa^{\times} \, \bigl(\kappa^{\perp}\bigr)^{-1} \Bigr)^2 \,\Bigr]
\end{equation*}
must be invertible. Notice that if $\kappa^{\,\perp}$ were symmetric its inverse would be well defined.
This topic will be addressed in the next section, devoted to the numerical results.
Concerning the entropy production rate in the LRF, this entails
\begin{equation}
\label{eq:entr_prod_mag}
    \nabla_{\mu}\mathcal{S}^{\mu} = \ldots + j_f^{\,i}\,\Bigl(\mathcal{C}_{f f'}^{\perp}\, \Xi^{il} + \mathcal{C}_{f f'}^{\parallel} \,\Hat{b}^{\,i} \Hat{b}^{\, l} + \mathcal{C}_{f f'}^{\times} \, \Hat{b}^{il}\Bigr)\,j_{f'}^{\,l} \geq 0\,,
\end{equation}
Provided that $\mathcal{C}_{f f'}^{\times}$ can be diagonalized the Hall current does not contribute to dissipation and one is left with
\be
\nabla_{\mu}\mathcal{S}^{\mu} = \ldots +
(j^{\,\perp})^i_f \,\mathcal{C}_{f f'}^{\,\perp}\, (j^{\,\perp})^i_\ff\,+ (j^{\,\parallel})^i_f \,\mathcal{C}_{f f'}^{\,\parallel}\, (j^{\,\parallel})^i_\ff\ge 0\,,\label{eq:entro_par+perp}
\ee
where
\begin{equation*}
 (j^{\,\perp})^i_f\equiv \Xi^{il}j^{\,l}_f\quad{\rm and}\quad
(j^{\,\parallel})^i_f\equiv \Hat{b}^i \Hat{b}^l\,j^{\,l}_f\,. 
\end{equation*}
Hence, positive eigenvalues of the $\mathcal{C}_{f f'}^{\,\perp}$ and $\mathcal{C}_{f f'}^{\,\parallel}$ matrices ensure a positive rate of entropy production. In Sec.~\ref{sec:num} we will verify that our RTA-BV calculation is actually consistent with this requirement.
\section{Numerical results}
\label{sec:num}
We now present our numerical results considering a conformal plasma composed of gluons and three species of massless quarks and antiquarks, fixing the free parameters in order to describe as closely as possible the experimental conditions in HIC's at RHIC and at the LHC. For this purpose, each set of figures refers to
\begin{itemize}
\item $s/n_B$ fixed: baryon number is conserved and the expansion of the fireball is approximately isentropic. Different values of entropy per baryon $s/n_B$ will be explored, the ratio being very high at the LHC and at top RHIC energies and lower in collisions with a higher baryon stopping (e.g. RHIC Beam-Energy Scan);
\item $n_Q/n_B=0.4\,|e|$: a non-zero electric charge of the fireball, conserved during its evolution, can only arise from the partial stopping of the incoming nuclear matter, hence $n_Q/n_B$ reflects the proton-to-nucleon ratio of the colliding nuclei;
\item $n_S=0$: the incoming nuclei do not carry strangeness and no net strangeness can be produced during the collision.
\end{itemize}
Starting from the EoS in Eq.~(\ref{eq:P-conformal}) and the thermodynamic relations in Eq.~(\ref{eq:therm_rel}), implying
\be
n_f=T^3\,\frac{g_f}{6}\,\left(\alpha_f+\frac{\alpha_f^3}{\pi^2}\right)
\ee
for the quark-flavor density and
\be
s= T^3\,\left[\frac{(4 g_g+7 \sum_f g_f)\pi^2}{90}+\sum_f\frac{g_f}{6}\,\alpha_f^2\right]
\ee
for the entropy density, one can fix the dimensionless quark chemical potentials $\{\alpha_f\}$ in order to fulfill the previous phenomenological constraints. Notice that, as shown in Eq.~(\ref{eq:set-conformal}), neglecting higher-order corrections in Kn and $\xi$, each fluid cell flows keeping $\alpha_f$ constant. The different sets of quark and charge chemical potentials employed in our calculations are reported in Table~\ref{tab:chem}. The three values of entropy per baryon are representative of collisions at intermediate RHIC energies ($s/n_B=50$), top RHIC energies ($s/n_B=300$)~\cite{Motta:2020cbr} and LHC energies ($s/n_B\to \infty$). Notice that, within the experimental uncertainties, data on matter-antimatter unbalance at mid-rapidity at the LHC are compatible with vanishing chemical potentials for all conserved charges~\cite{ALICE:2023ulv}. This is the situation of relevance to describe also the QGP filling the early universe, as inferred from the very large photon-to-baryon ratio of order $10^9$~\cite{Planck:2018vyg}.
\begin{table}[!hbt]
\centering
\begin{tabular}{llllll}
\hline\noalign{\smallskip}
 & $\alpha_u$ & $\alpha_d$ & $\alpha_{\cal{B}}$ & $\alpha_Q$ & $\alpha_S$ \\
\noalign{\smallskip}\hline\noalign{\smallskip}
$s/n_B = 50$ & 0.585 & 0.662 & 1.909 & -0.254 & 0.662 \\
$s/n_B = 300$ & 0.097 & 0.111 & 0.319 & -0.046 & 0.111 \\
$s/n_B \rightarrow \infty$ & 0.0 & 0.0 & 0.0 & 0.0 & 0.0  \\
\noalign{\smallskip}\hline
\end{tabular}
\caption{Dimensionless chemical potentials associated with quarks and conserved charges for different values of entropy per baryon. The strange-quark chemical potential $\alpha_s$ is omitted since it identically vanishes for any value of $s/n_B$, due to the assumption of zero strangeness in the system.}
\label{tab:chem} 
\end{table}

For a conformal plasma the temperature is the only physical scale of the problem and all quantities can be made dimensionless by rescaling them by proper powers of the temperature. Also the magnetic field, exploiting Eq.~(\ref{eq:beta-inv}), can be conveniently expressed as
\begin{equation}
B/T^2=\sqrt{\,2\,(P/T^4)\,\beta_V^{-1}}\,.
\end{equation}
Accordingly in the following, at fixed $\{\alpha_f\}$, we display results for the dimensionless flavor-diffusion matrices $\kappa_{ff'}/T^2$ and charge conductivities $\sigma_q/T$ plotted as functions of the actual scaling variable governing the behavior of the system: the plasma inverse beta-value $\beta_V^{-1}$.
Furthermore, our results refer to a fixed value of the dimensionless relaxation time
\begin{equation}
    \widetilde{\tau}_R\equiv \tau_R\, T\,.
\end{equation}
At zero chemical potential the latter is related to the specific shear-viscosity $\eta/s$ by Eq.~(\ref{eq:BOE}), which for $\eta/s\approx 0.2$ gives a result of order one. Hence, in the following we set $\widetilde\tau_R=1$ also for the case of non-zero quark density.

\begin{figure*}[!hbt]
    \centering
    \includegraphics[width=0.32\textwidth] {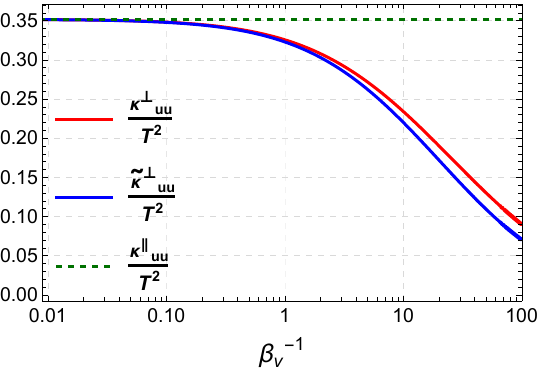} 
    \includegraphics[width=0.328\textwidth] {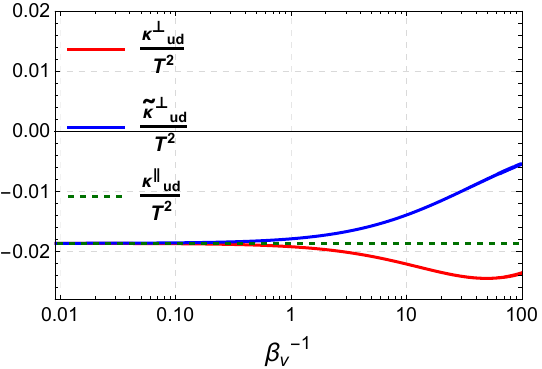}
    \includegraphics[width=0.332\textwidth] {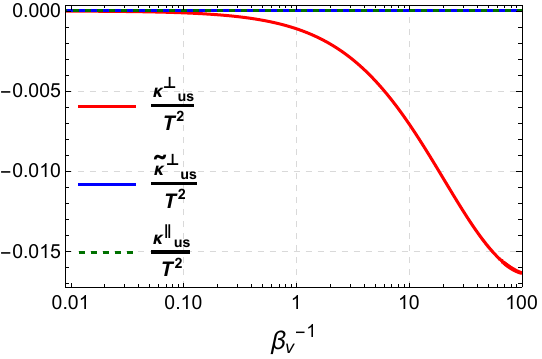} \\
    \includegraphics[width=0.33\textwidth] {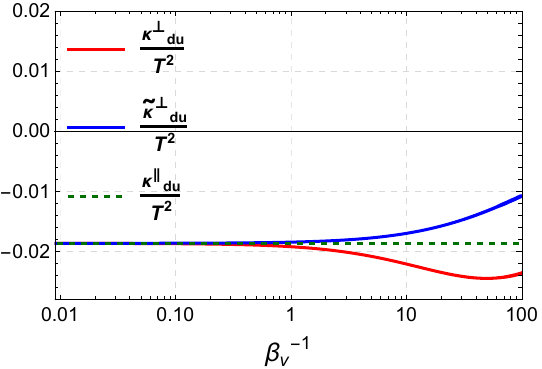} 
    \includegraphics[width=0.319\textwidth] {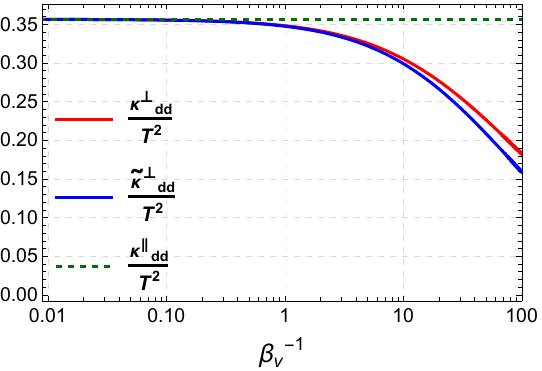}
    \includegraphics[width=0.327\textwidth] {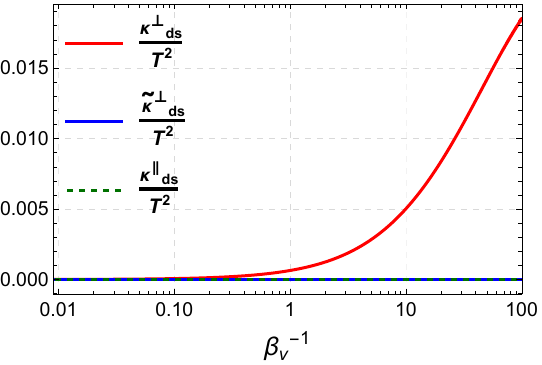}\\
    \includegraphics[width=0.333\textwidth] {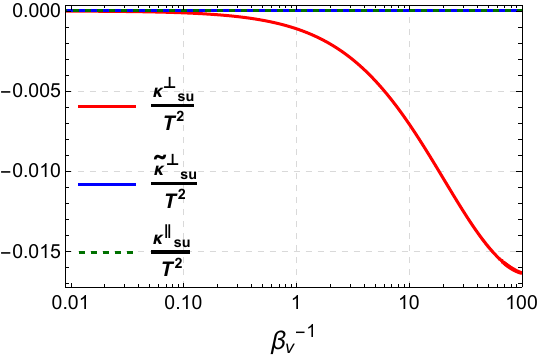} 
    \includegraphics[width=0.33\textwidth] {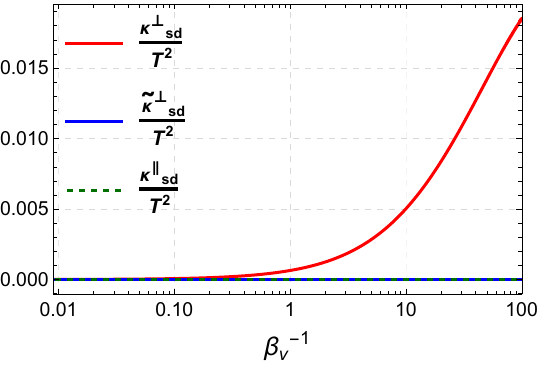}
    \includegraphics[width=0.325\textwidth] {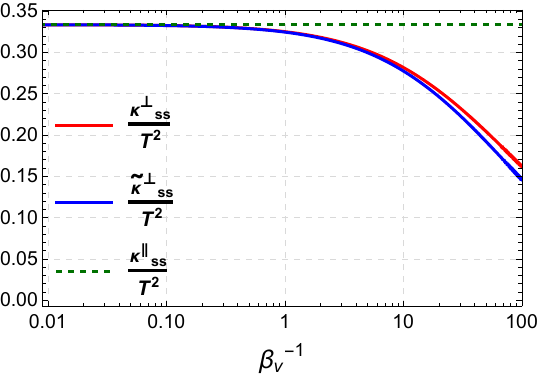}   
    \caption{Matrix elements in flavor space of the transverse component of the quark-diffusion tensor as a function of the inverse plasma beta-value $\beta_V^{-1}$. Here we consider the case $s/n_B = 50$ (intermediate RHIC energies), corresponding to the dimensionless chemical potentials in Table~\ref{tab:chem}. The dimensionless relaxation time is set to $\widetilde\tau_R=1$. Continuous red curves: full result given by Eq.~(\ref{eq:system}). Continuous blue curves: results from Eq.~(\ref{eq:kort}), without the mixing of the transverse and Hall responses. Dashed green curves: longitudinal response, unaffected by the magnetic field.}
    \label{fig:kp_Bv}
\end{figure*}
\begin{figure*}[!hbt]
    \centering
    \includegraphics[width=0.32\textwidth] {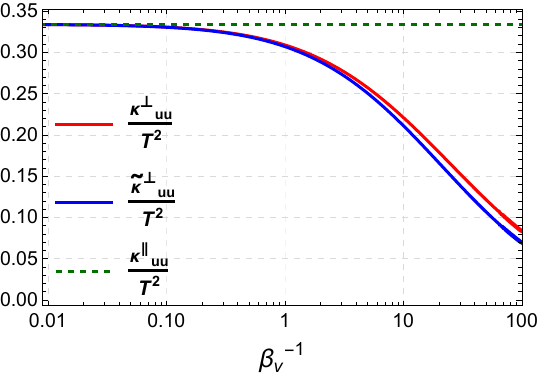} 
    \includegraphics[width=0.329\textwidth] {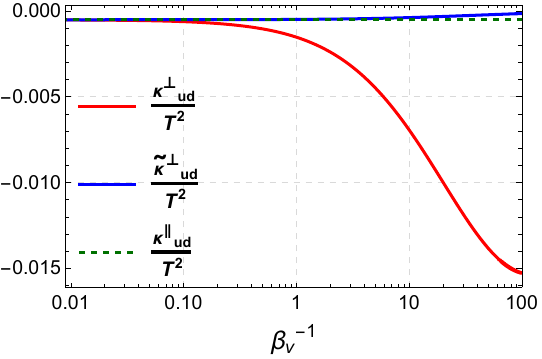}
    \includegraphics[width=0.332\textwidth] {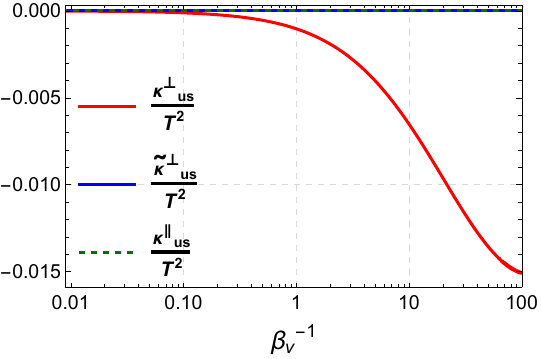} \\
    \includegraphics[width=0.332\textwidth] {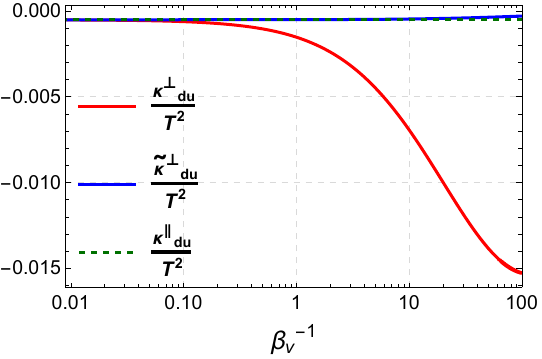} 
    \includegraphics[width=0.324\textwidth] {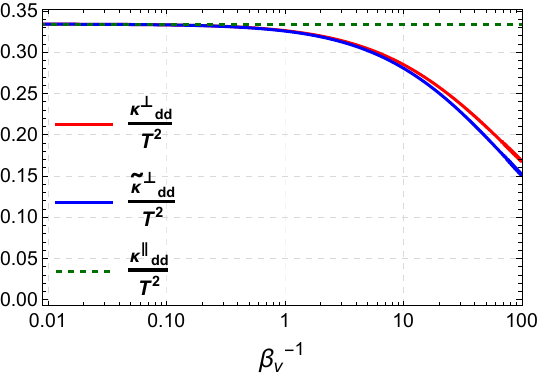}
    \includegraphics[width=0.334\textwidth] {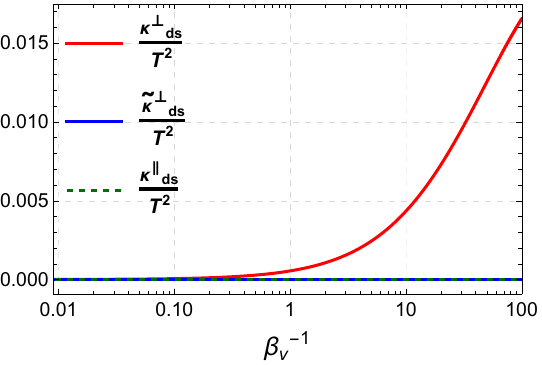}\\
    \includegraphics[width=0.335\textwidth] {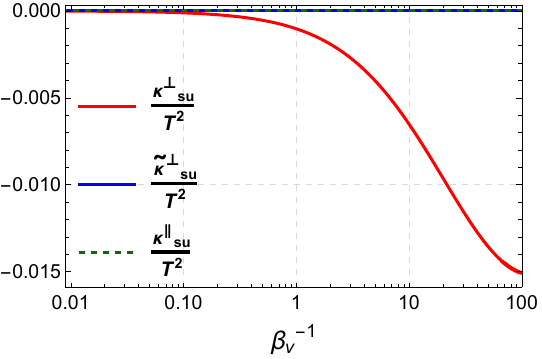} 
    \includegraphics[width=0.331\textwidth] {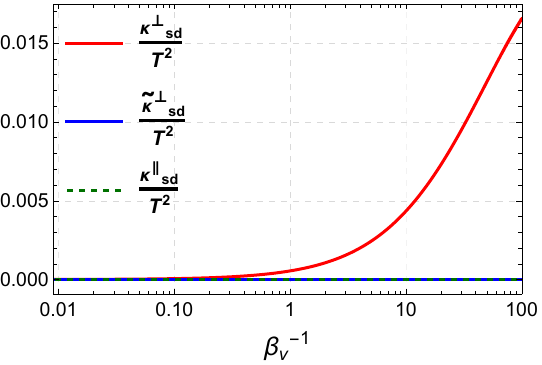}
    \includegraphics[width=0.325\textwidth] {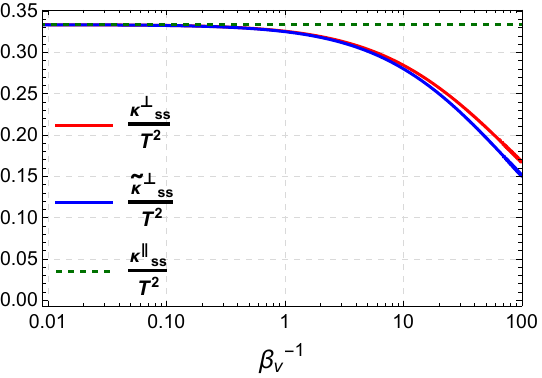} 
    \caption{The same as in Fig.~\ref{fig:kp_Bv}, but for $s/n_B = 300$ (top RHIC energies).}
    \label{fig:kp_Bv_re}
\end{figure*}
\begin{figure*}[!hbt]
    \centering
    \includegraphics[width=0.32\textwidth] {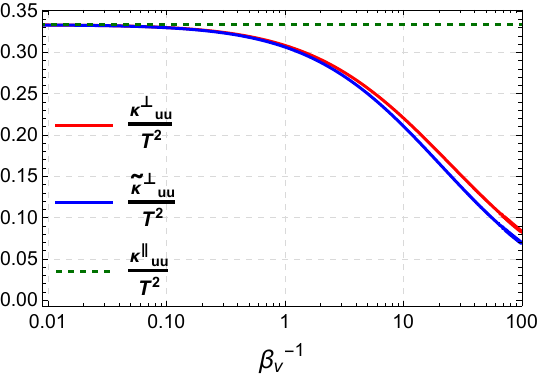} 
    \includegraphics[width=0.331\textwidth] {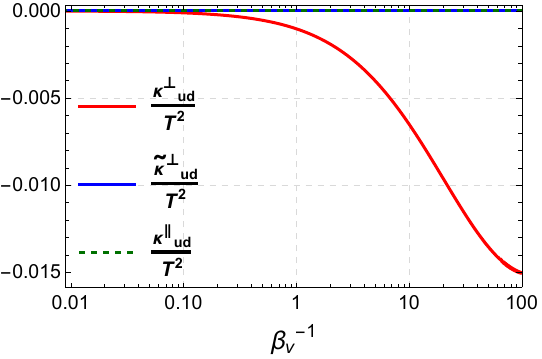}
    \includegraphics[width=0.331\textwidth] {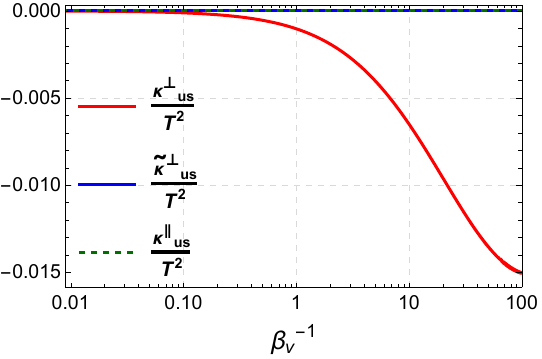} \\
    \includegraphics[width=0.333\textwidth] {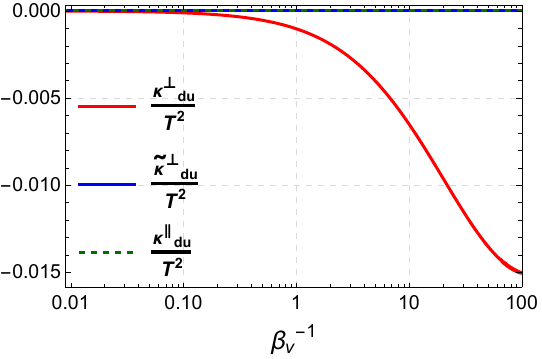} 
    \includegraphics[width=0.324\textwidth] {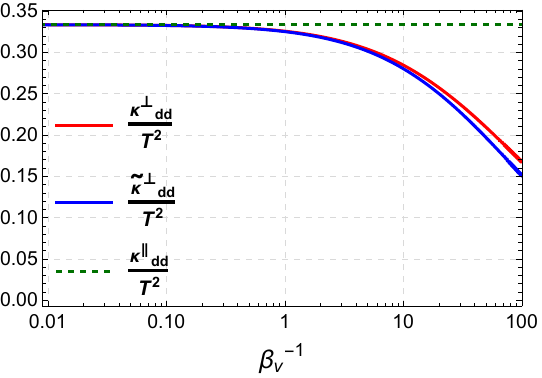}
    \includegraphics[width=0.333\textwidth] {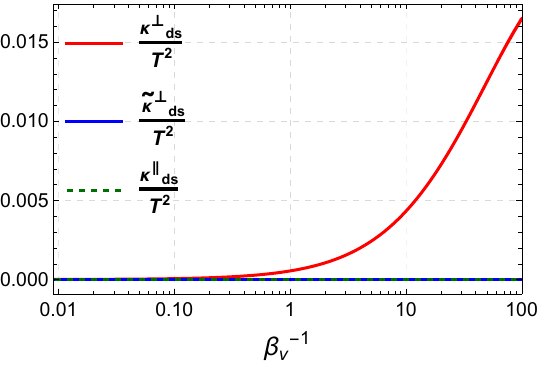}\\
    \includegraphics[width=0.334\textwidth] {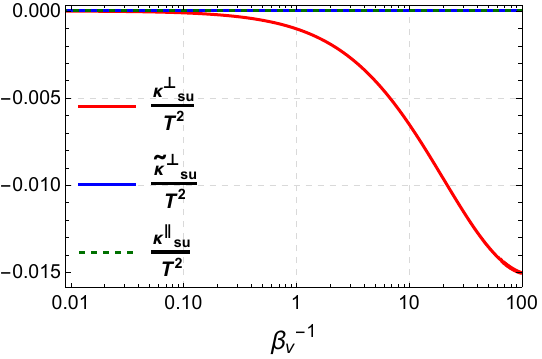} 
    \includegraphics[width=0.331\textwidth] {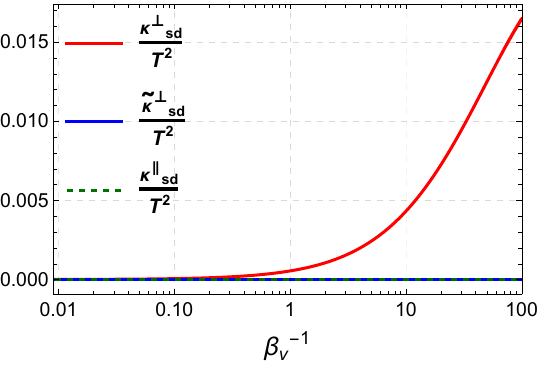}
    \includegraphics[width=0.326\textwidth] {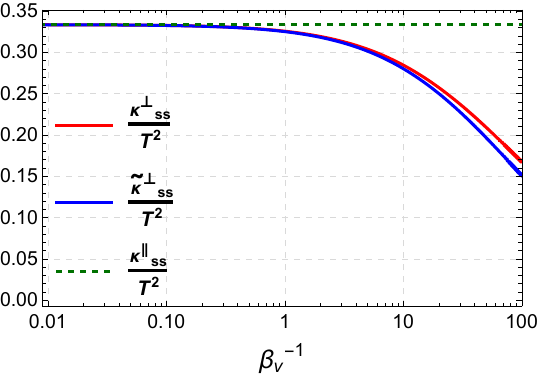}  
    \caption{The same as in Fig.~\ref{fig:kp_Bv}, but for $s/n_B \rightarrow \infty$ (top LHC energies and early universe).}
    \label{fig:kp_Bv_re3}
\end{figure*}
We start considering the transverse component of the quark-diffusion tensor. Results for the cases of entropy per baryon $s/n_B=50$, $s/n_B=300$ and $s/n_B\to\infty$ are shown in Figs.~\ref{fig:kp_Bv},~\ref{fig:kp_Bv_re} and~\ref{fig:kp_Bv_re3}, where in the different panels we plot all its matrix elements $\kappa^{\,\perp}_{ff'}$ in flavor space (red curves). 
It turns out from the numerical calculation that the full matrix $\kappa^{\,\perp}$ is almost diagonal in flavor space and that off-diagonal matrix elements are subdominant with respect to diagonal ones; furthermore, all matrix elements are non-vanishing for large enough magnetization.\\ 
As long as the plasma is weakly magnetized all curves for $\kappa^\perp_{ff'}$ are flat, coinciding with the longitudinal response $\kappa^{||}_{ff'}$ (dashed green curves), which remains independent of $\beta_V^{-1}$ for any value of the magnetic field, consistently with Eq.~(\ref{eq:kpar}). 
Hence for weak magnetic field one has
$\kappa^\perp_{ff'}\!=\!\kappa^{||}_{ff'}\!=\!\kappa_{ff'}$. As it follows from Eq.~(\ref{eq:kappa-weak}) the latter is symmetric~\cite{Fotakis:2019nbq,Gavassino:2023qnw} and all its off-diagonal elements involving an $s$-index identically vanish, since we are considering the case of a medium with zero net strange-quark density. 
Considering the case -- addressed in Fig.~\ref{fig:kp_Bv_re3} -- in which {\em all} quark chemical potentials vanish, one can see that for weak magnetization $\kappa^{\,\perp}_{f\ff}$, coinciding in this limit with $\kappa_{f\ff}$, acquires a purely diagonal structure, consistently with Eq.~(\ref{eq:kappa-weak}).\\
We also display in the different panels the matrix elements of $\widetilde\kappa^{\,\perp}$, for which one neglects the contribution from the Lorentz force to the fluid acceleration. Notice how differences between $\kappa^{\perp}$ and $\widetilde\kappa^{\,\perp}$ are small for the diagonal matrix elements, but sizable for the off-diagonal ones.
All off-diagonal matrix elements of $\widetilde\kappa^{\,\perp}$ involving an $s$ index identically vanish, consistently with Eqs.~(\ref{eq:kort}) and~(\ref{eq:Ap-Am}).\\
Independently of the value of $s/n_B$, the major common feature of the diagonal matrix elements of $\kappa^{\,\perp}$ is their monotonous decrease as the plasma magnetization gets larger. This is due to the Larmor bending of the trajectories of charged particles in the plane orthogonal to the magnetic field, responsible for the denominator in the integrand in Eq.~(\ref{eq:kort}).

\begin{figure*}[!hbt]
    \centering
    \includegraphics[width=0.32\textwidth] {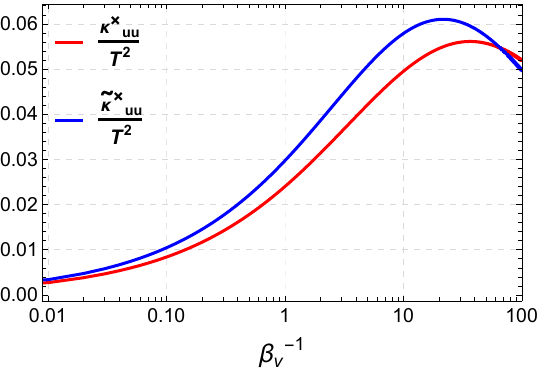} 
    \includegraphics[width=0.33\textwidth] {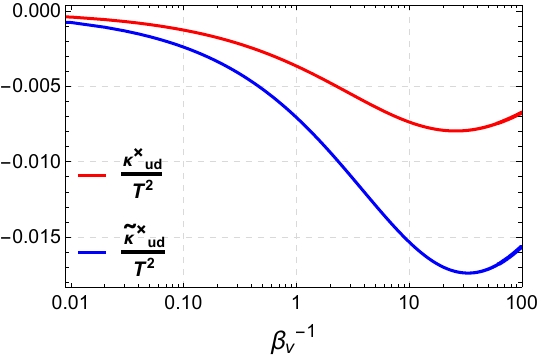}
    \includegraphics[width=0.325\textwidth] {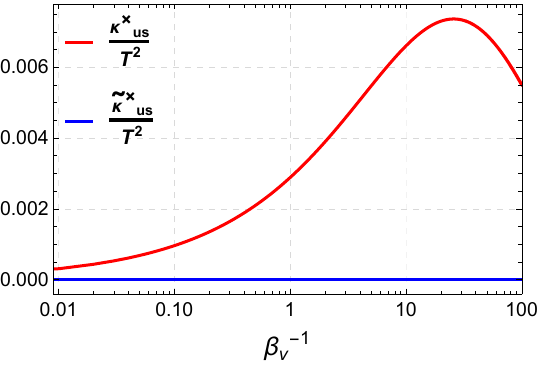} \\
    \includegraphics[width=0.33\textwidth] {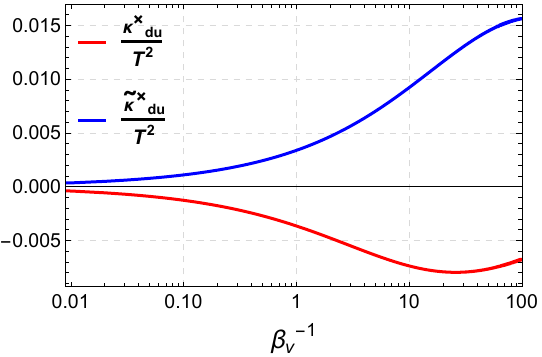} 
    \includegraphics[width=0.323\textwidth] {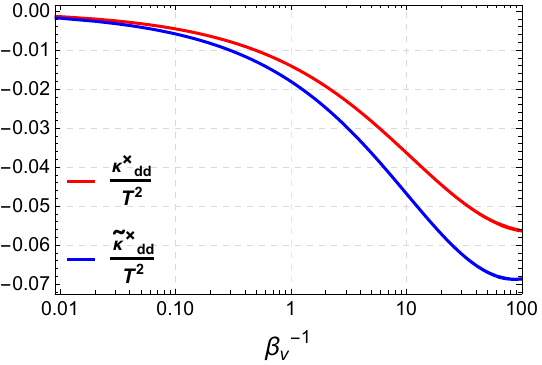}
    \includegraphics[width=0.327\textwidth] {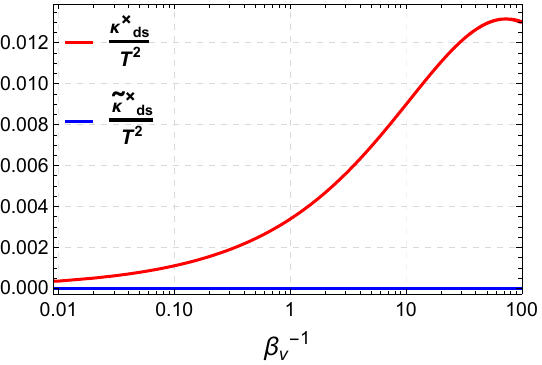}\\
    \includegraphics[width=0.328\textwidth] {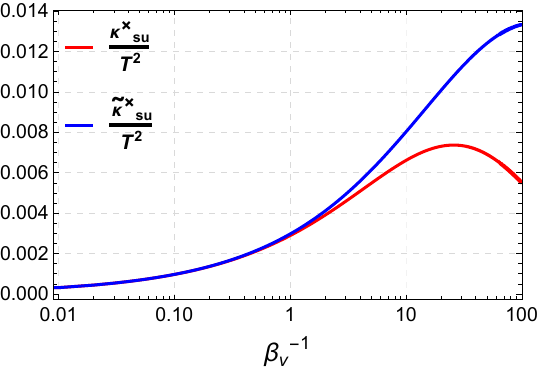} 
    \includegraphics[width=0.328\textwidth] {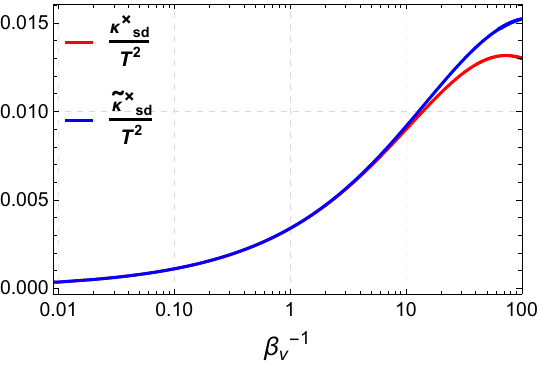}
    \includegraphics[width=0.334\textwidth] {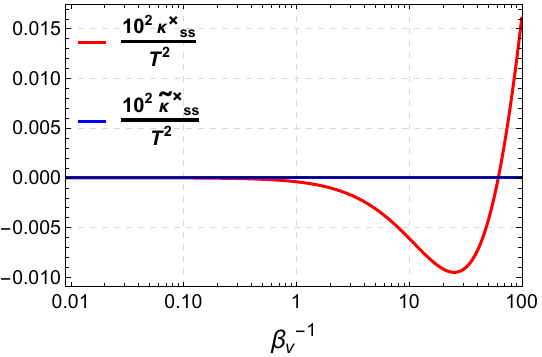} 
    \caption{Matrix elements in flavor space of the Hall component of the quark-diffusion tensor as a function of the inverse plasma beta-value $\beta_V^{-1}$. Here we consider the case $s/n_B = 50$ (intermediate RHIC energies), corresponding to the dimensionless chemical potentials in Table~\ref{tab:chem}. The dimensionless relaxation time is set to $\widetilde\tau_R=1$. Continuous red curves: full result given by Eq.~(\ref{eq:system}). Continuous blue curves: results from Eq.~(\ref{eq:kx}), without the mixing of the transverse and Hall responses.}
    \label{fig:kx_Bv}
\end{figure*}
\begin{figure*}[!hbt]
    \centering
    \includegraphics[width=0.32\textwidth] {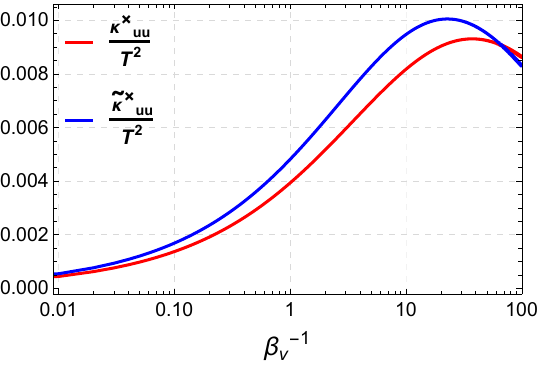} 
    \includegraphics[width=0.334\textwidth] {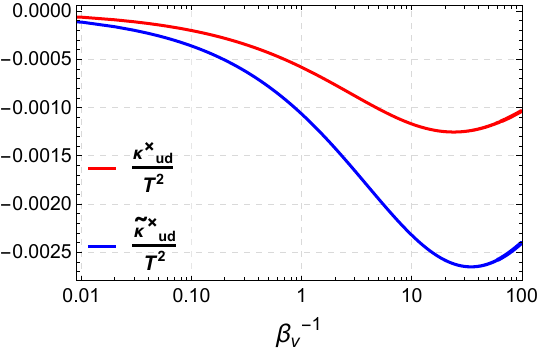}
    \includegraphics[width=0.325\textwidth] {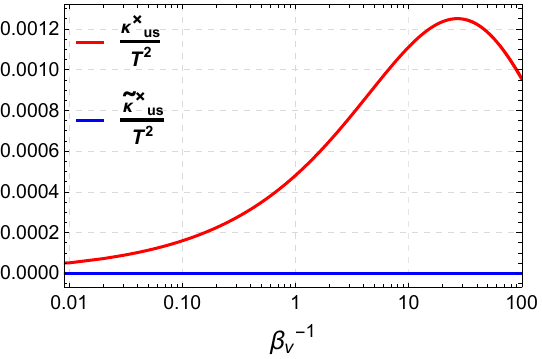} \\
    \includegraphics[width=0.33\textwidth] {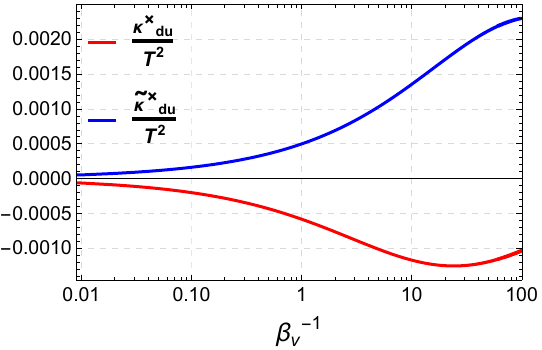} 
    \includegraphics[width=0.328\textwidth] {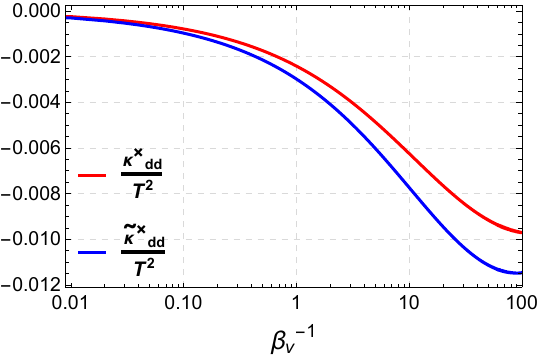}
    \includegraphics[width=0.329\textwidth] {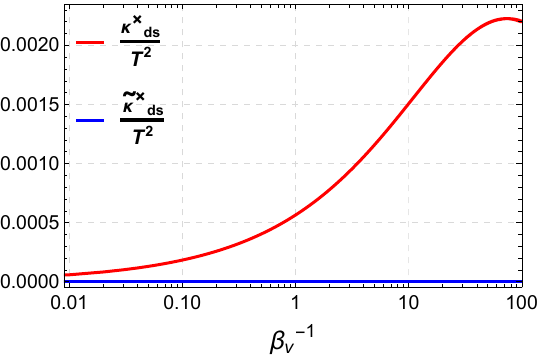}\\
    \includegraphics[width=0.328\textwidth] {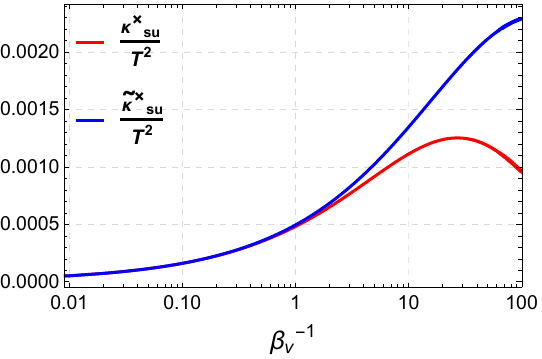} 
    \includegraphics[width=0.33\textwidth] {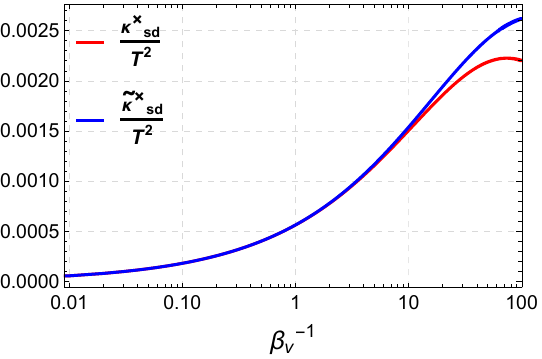}
    \includegraphics[width=0.334\textwidth] {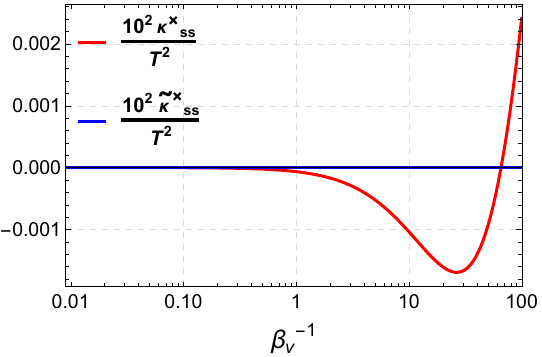}   
    \caption{The same as in Fig.~\ref{fig:kx_Bv}, but for $s/n_B = 300$ (top RHIC energies).}
    \label{fig:kx_Bv_re}
\end{figure*}
We now move to the Hall component of the quark diffusion tensor, whose matrix elements in flavor space are plotted in Figs.~\ref{fig:kx_Bv} and~\ref{fig:kx_Bv_re} for the cases $s/n_B=50$ and $s/n_B=300$, respectively. Red curves refer to the full result,  $\kappa^{\times}_{ff'}$, arising from the solution of the system of coupled equations~(\ref{eq:system}). Blue curves come from Eq.~(\ref{eq:kx}), where $\widetilde\kappa^{\,\times}_{ff'}$ is again obtained by neglecting the Lorentz force in the Euler equation for the fluid acceleration. Also $\kappa^{\,\times}$ turns out to be symmetric in the flavor indices. Diagonal matrix elements of $\kappa^{\,\times}$ are much smaller than the ones of $\kappa^{\,\perp}$.
All matrix elements of $\kappa^{\,\times}$ rapidly decrease going to larger values of $s/n_B$; in particular, $\kappa^{\,\times}$ identically vanishes in the $s/n_B\to\infty$ limit, since a non-zero Hall current can develop only in the presence of a particle-antiparticle imbalance.
Notice that, due to the vanishing net strange-quark density in HIC's, $\widetilde\kappa^{\,\times}_{ss}=0$ and  $\kappa^{\times}_{ss}$ is in any case very small, a non-zero result arising only from the mixing introduced by the system in Eq.~(\ref{eq:system}).
For the same reason, also off-diagonal matrix elements of the form $\widetilde\kappa^{\,\times}_{fs}$ identically vanish, as it follows from Eqs.~(\ref{eq:kx}) and~(\ref{eq:kx-integrand}).

\begin{figure*}[!hbt]
    \centering
    \includegraphics[width=0.45\textwidth] {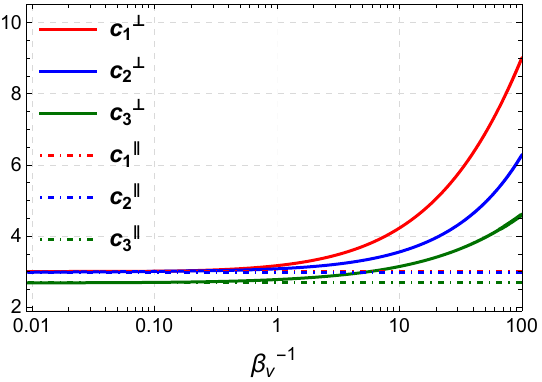}
    \includegraphics[width=0.455\textwidth] {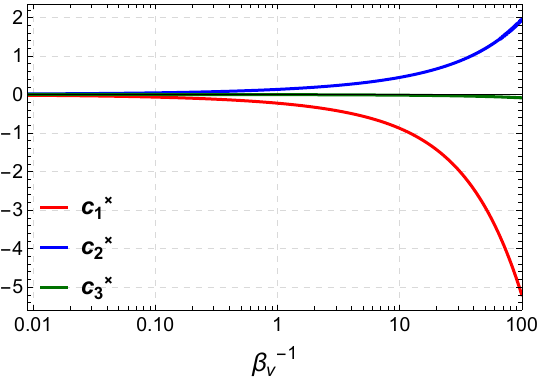} 
    \caption{Eigenvalues of the dimensionless matrices $T^2\,\mathcal{C}^{\,\perp}$, $T^2\,\mathcal{C}^{\,\parallel}$ (left panel) and $T^2\,\mathcal{C}^{\,\times}$ (right panel) for the case $s/n_B=50$. All eigenvalues of the $T^2\,\mathcal{C}^{\,\perp}$, $T^2\,\mathcal{C}^{\,\parallel}$ matrices are positive, entailing a positive contribution to entropy production. All eigenvalues of $T^2\,\mathcal{C}^{\,\times}$ are distinct, implying a vanishing contribution to entropy production from the Hall flavor currents.}
    \label{fig:Aeigen}
\end{figure*}
\begin{figure*}[!hbt]
    \centering
    \includegraphics[width=0.45\textwidth] {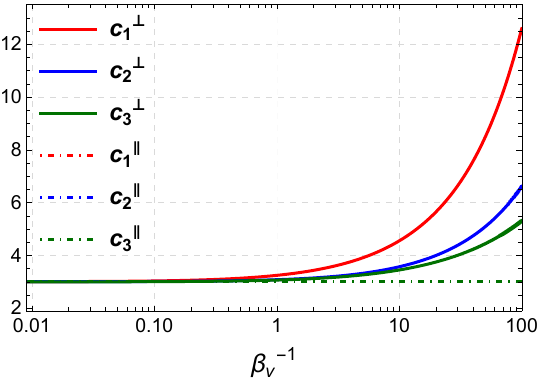} 
    \includegraphics[width=0.465\textwidth] {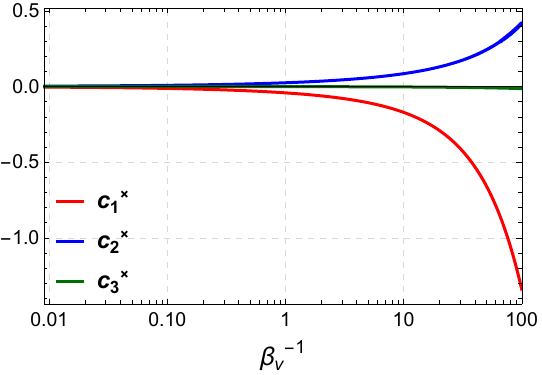} 
    \caption{The same as in Fig.~\ref{fig:Aeigen}, but for $s/n_B=300$.}
    \label{fig:Aeigen_re}
\end{figure*}
\begin{figure*}[!hbt]
    \centering
    \includegraphics[width=0.45\textwidth] {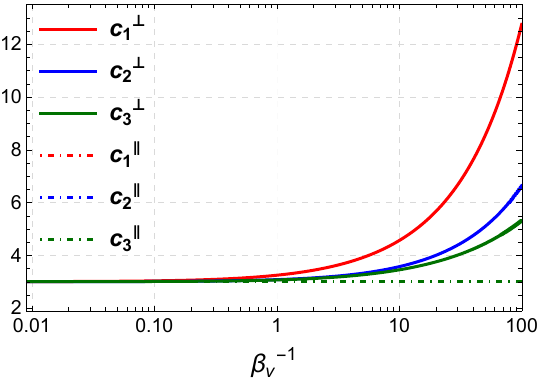} 
    \includegraphics[width=0.475\textwidth] {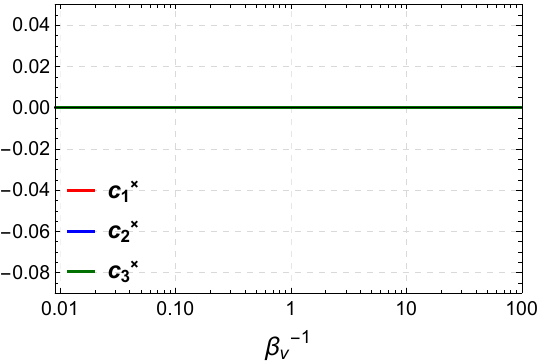} 
    \caption{The same as in Fig.~\ref{fig:Aeigen}, but for $s/n_B\to\infty$. For zero net-quark density both $\kappa^{\,\times}$ and hence $C^{\,\times}$ identically vanish and accordingly the eigenvalues of the latter.}
    \label{fig:Aeigen_re3}
\end{figure*}
We now address the issue of entropy production. For this purpose, in Figs.~\ref{fig:Aeigen},~\ref{fig:Aeigen_re} and~\ref{fig:Aeigen_re3} we show the eigenvalues of the $\mathcal{C}^{\,\parallel}$, $\mathcal{C}^{\,\perp}$ and $\mathcal{C}^{\,\times}$ matrices defined in Eqs.~(\ref{eq:Cpar}),~(\ref{eq:Cperp}) and~(\ref{eq:Ccross}). From the left panels one can conclude that all the eigenvalues of $\mathcal{C}^{\,\parallel}$, $\mathcal{C}^{\,\perp}$ are positive, entailing -- according to Eq.~(\ref{eq:entro_par+perp}) -- a positive contribution to entropy production from the longitudinal and transverse induced flavor currents. At the same time, having three distinct eigenvalues for $\mathcal{C}^{\,\times}$ independently of the value of $\beta_V^{\,-1}$ ensures that $\mathcal{C}^{\,\times}$ can be diagonalized and hence the Hall currents never contribute to entropy production. This is in agreement with what was independently found in Ref.~\cite{Ammon:2020rvg} within a holographic approach. Notice in Fig.~\ref{fig:Aeigen_re3} that all the eigenvalues of $C^{\,\times}$ vanish, since in the $s/n_B\to\infty$ limit one has $\kappa^{\,\times}=C^{\,\times}=0$: in this case the absence of entropy production from the Hall current trivially follows from the fact that that the latter is identically zero in the absence of a quark-antiquark imbalance.

\begin{figure*}[!hbt]
    \centering
    \includegraphics[width=0.46\textwidth] {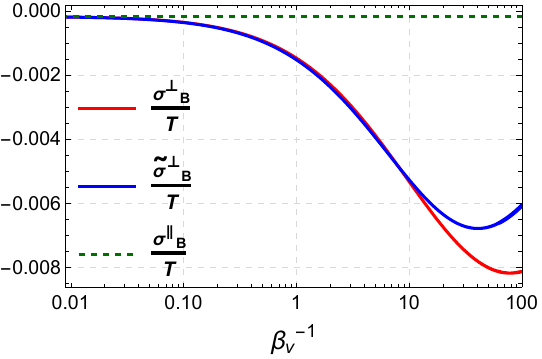} 
    \includegraphics[width=0.451\textwidth] {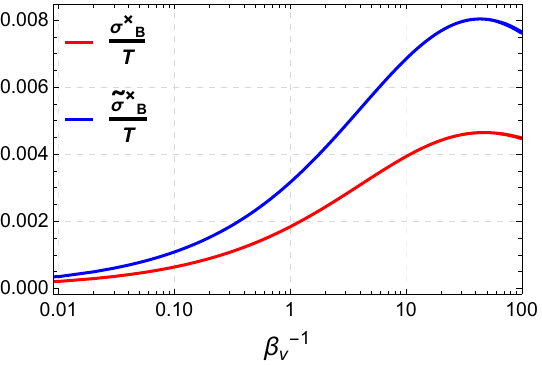}\\
    \includegraphics[width=0.45\textwidth] {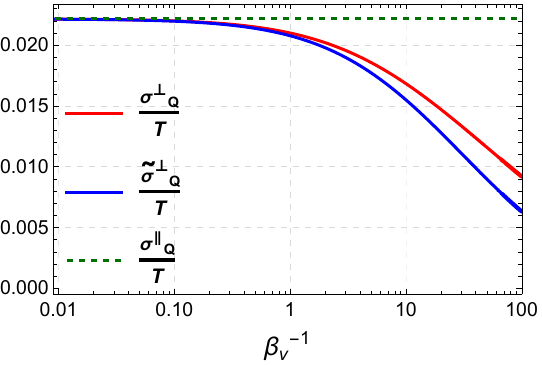} 
    \includegraphics[width=0.45\textwidth] {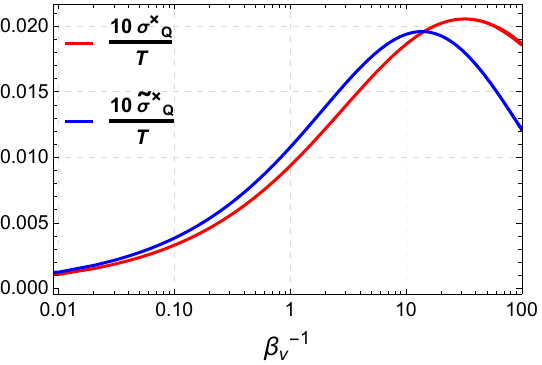}\\
    \includegraphics[width=0.45\textwidth] {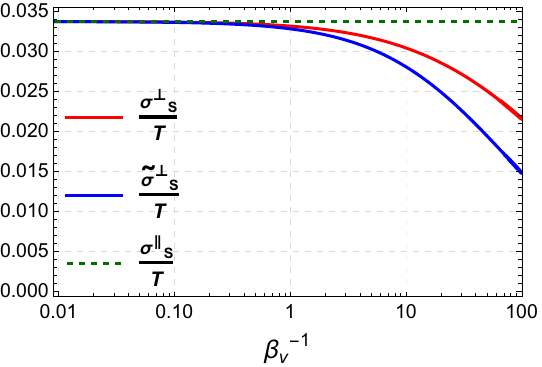} 
    \includegraphics[width=0.46\textwidth] {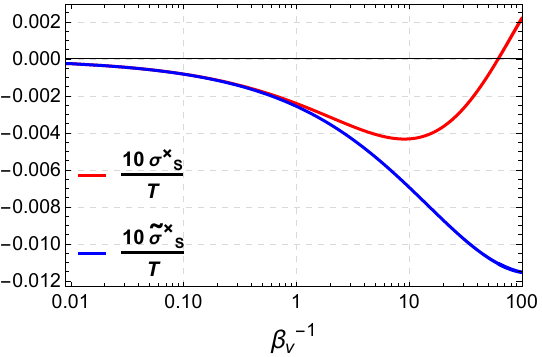}
    \caption{Transverse, longitudinal (left panels) and Hall (right panels) components of the conductivities associated with the conserved charges -- baryon number ${\cal B}$ (upper row), electric charge $Q$ (central row) and strangeness $S$ (bottom row) -- plotted as functions of the inverse beta-value.
    Displayed curves refer to a fireball with entropy per baryon $s/n_B = 50$.}
    \label{fig:cond}
\end{figure*}
\begin{figure*}[!hbt]
    \centering
    \includegraphics[width=0.451\textwidth] {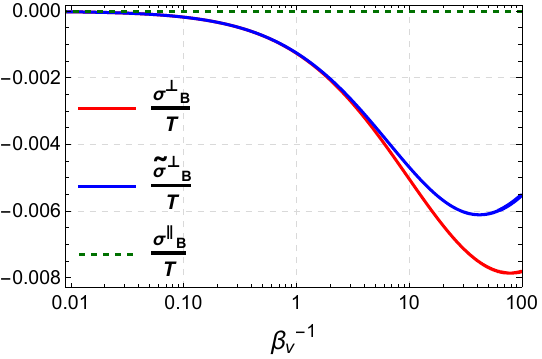} 
    \includegraphics[width=0.454\textwidth] {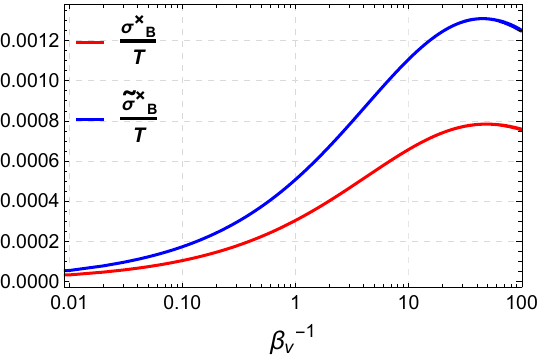}\\
    \includegraphics[width=0.45\textwidth] {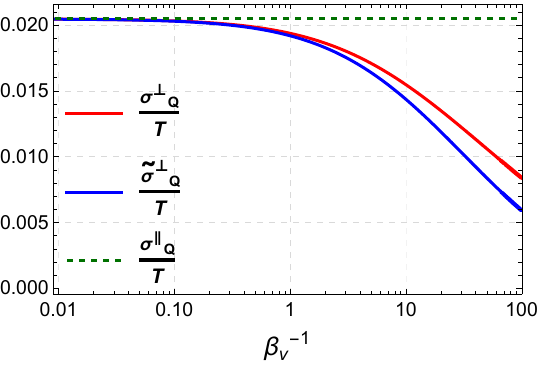} 
    \includegraphics[width=0.46\textwidth] {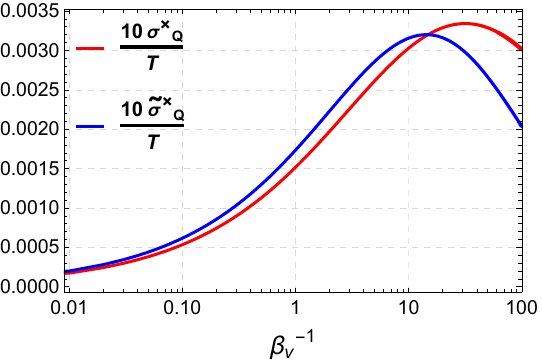}\\
    \includegraphics[width=0.45\textwidth] {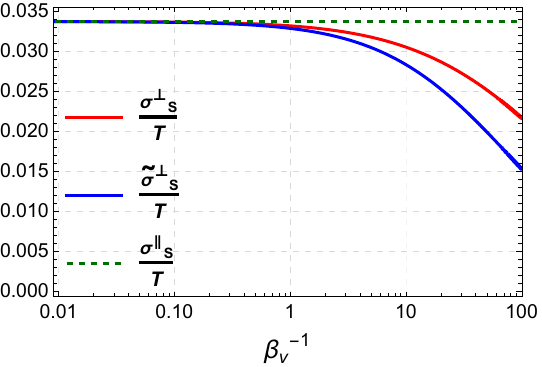} 
    \includegraphics[width=0.47\textwidth] {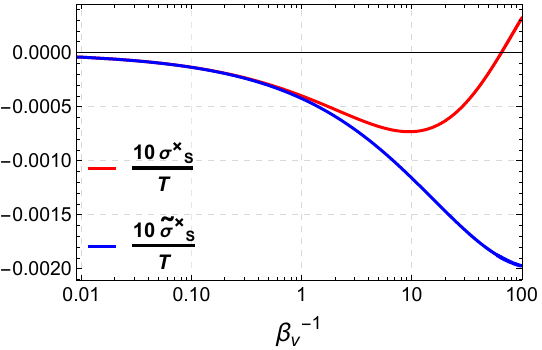}
    \caption{The same as in Fig.~\ref{fig:cond}, but for $s/n_B = 300$.}
    \label{fig:cond_re}
\end{figure*}
\begin{figure*}[!hbt]
    \centering
    \includegraphics[width=0.451\textwidth] {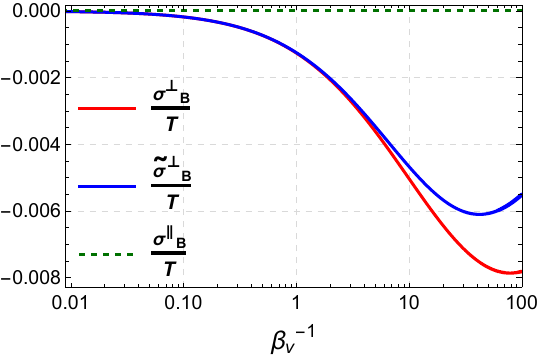} 
    \includegraphics[width=0.454\textwidth] {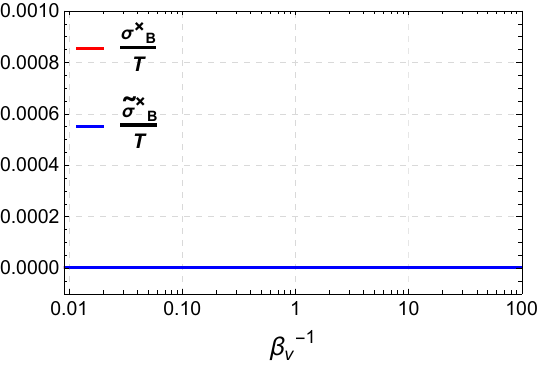}\\
    \includegraphics[width=0.45\textwidth] {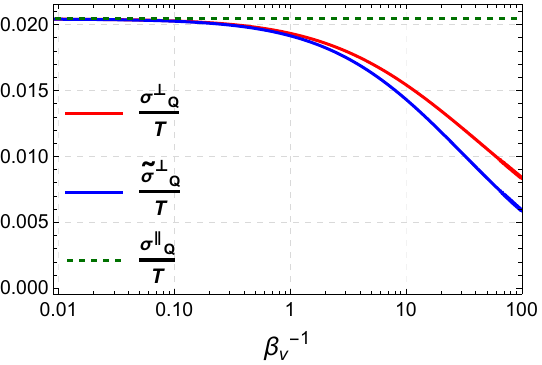} 
    \includegraphics[width=0.46\textwidth] {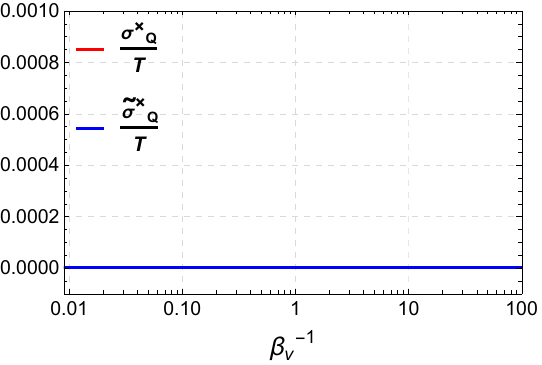}\\
    \includegraphics[width=0.45\textwidth] {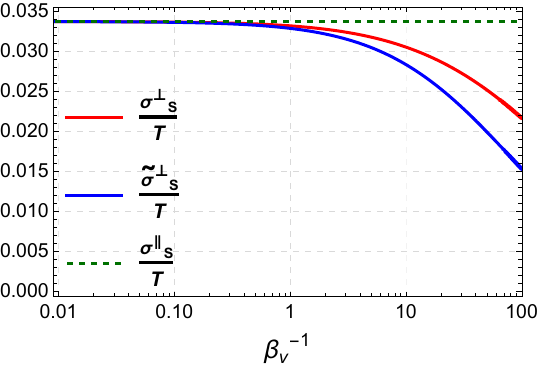} 
    \includegraphics[width=0.47\textwidth] {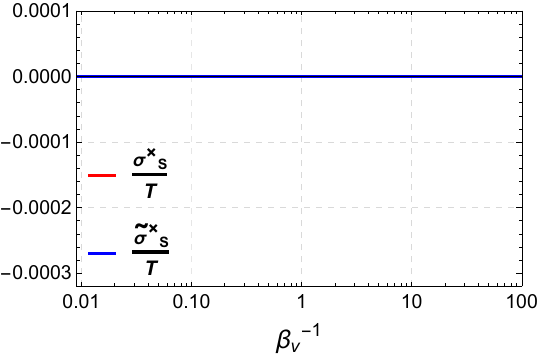}
    \caption{The same as in Fig.~\ref{fig:cond}, but for $s/n_B \rightarrow \infty$. For zero net-quark density the Hall conductivities (right panels) identically vanish.}
    \label{fig:cond_re3}
\end{figure*}
In Figs.~\ref{fig:cond},~\ref{fig:cond_re} and~\ref{fig:cond_re3} we display the transverse, longitudinal and Hall components of the conductivity tensor associated with the conserved charges of the system: baryon number ${\cal B}$, electric charge $Q$ and strangeness $S$. Their link with the flavor-diffusion tensor follows from the generalization of Eq.~(\ref{eq:WF0}) to the case of a strongly-magnetized plasma: 
\be
\sigma_q^{\, a}\!=\!\sum_{f,\ff}{\cal M}_{qf}\,\frac{\kappa_{ff'}^{\, a}}{T}\,\varmathbb{Q}_{f'}|e|=\!\sum_{f,\ff} q_f\,\frac{\kappa_{ff'}^{\, a}}{T}\,\varmathbb{Q}_{f'}|e|\quad(a=\perp,\parallel,\times)\label{eq:WF-gen}\,.
\ee
Considering Fig.~\ref{fig:cond}, referring to the case $s/n_B\!=\!50$, one can see that, even for weak magnetization, the longitudinal and transverse baryon-number conductivities are slightly negative. This comes from the fact that, when a small quark-antiquark imbalance is present in the system due to the partial stopping of the incoming nuclei, the latter, according to Table~\ref{tab:chem}, is dominated by an excess of $d$ quarks, carrying negative electric charge (hence moving in the opposite direction with respect to the local electric field), but positive baryon number. For all values of $s/n_B$ the behavior of the transverse baryon conductivity for large magnetization can be understood observing that the generalized Wiedemann-Franz law in Eq.~(\ref{eq:WF-gen}) is dominated by the diagonal components of $\kappa^{\,\perp}_{f\ff}$, such that
{\setlength\arraycolsep{1pt}
\begin{eqnarray*}
\sigma_B^{\,\perp}&\approx& \frac{1}{3}\left[ \kappa^{\,\perp}_{uu}(\frac{2}{3}|e|)+\kappa^{\,\perp}_{dd}(-\frac{1}{3}|e|)+\kappa^{\,\perp}_{ss}(-\frac{1}{3}|e|)\right] \approx\\
{}&\approx&\frac{2}{9}|e|\left( \kappa^{\,\perp}_{uu}-\kappa^{\,\perp}_{dd}\right)<0
\end{eqnarray*}}
where in the last line we exploited the fact that, for diagonal matrix elements in flavor space, one has $\kappa^{\,\perp}\!\approx\!\widetilde\kappa^{\,\perp}$, $\widetilde\kappa^{\,\perp}_{dd}\!\approx\! \widetilde\kappa^{\,\perp}_{ss}$ and $\widetilde\kappa^{\,\perp}_{uu}<\widetilde\kappa^{\,\perp}_{dd}$ due to stronger suppression in Eq.~(\ref{eq:kort}) arising from the larger electric charge of $u$ quarks. Considering now the electric conductivity, both its transverse and longitudinal components are clearly positive, but the transverse one, for large magnetization, is suppressed due to the Larmor bending of the trajectories of charged particles, responsible for the denominator in the integrand in Eq.~(\ref{eq:kort}). Finally, for what concerns the strangeness conductivity, both $\sigma_{S}^{\,\perp}$ and $\sigma_{S}^{\,\parallel}$ are positive, since both the electric charge and the strangeness carried by $s$ quarks are negative. Also in this case the decrease of $\sigma_{S}^{\,\perp}$ for large values of $\beta_V^{\,-1}$ is due the bending of the particle trajectories in the plane orthogonal to the magnetic field. 
Regarding the Hall conductivities $\sigma^{\,\times}_q$, as expected, they all decrease going from small to large values of entropy per baryon, until identically vanishing for $s/n_B\to\infty$, as can be seen in Fig~\ref{fig:cond_re3}. The peculiar behavior of $\sigma_S^{\,\times}$ for large magnetization can be understood taking into account that, $\kappa_{ss}^{\,\times}$ being very small as compared to the other matrix elements, one approximately has
\begin{equation*}
  \frac{\sigma_S^{\,\times}}{T}\approx-\frac{2}{3}|e|\,\frac{\kappa_{su}^{\,\times}}{T^2}+ \frac{1}{3}|e|\,\frac{\kappa_{sd}^{\,\times}}{T^2}\,, 
\end{equation*}
leading to a change of sign for large values of $\beta_V^{\,-1}$.

\begin{figure*}[!hbt]
    \centering
    \includegraphics[width=0.45\textwidth] {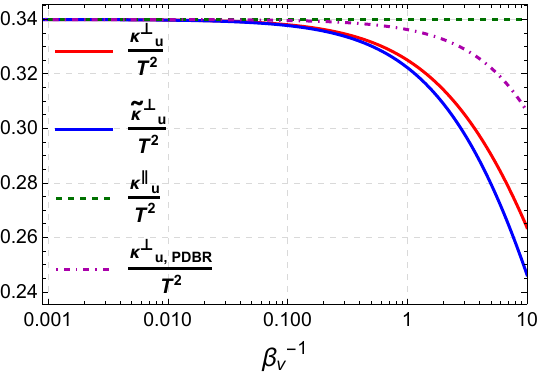} 
    \includegraphics[width=0.45\textwidth] {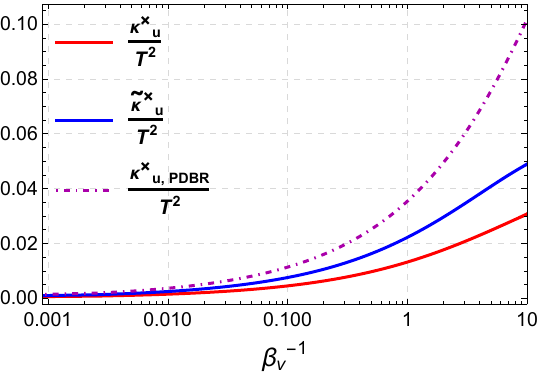}\\
    \includegraphics[width=0.45\textwidth] {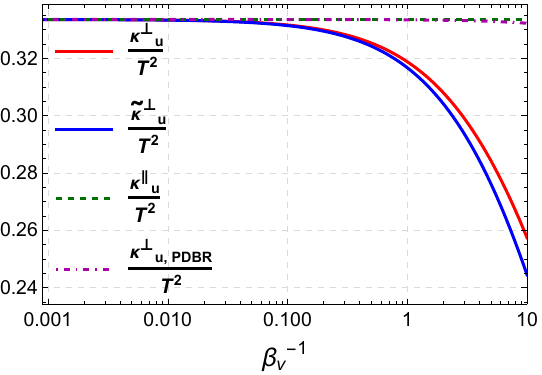} 
    \includegraphics[width=0.46\textwidth] {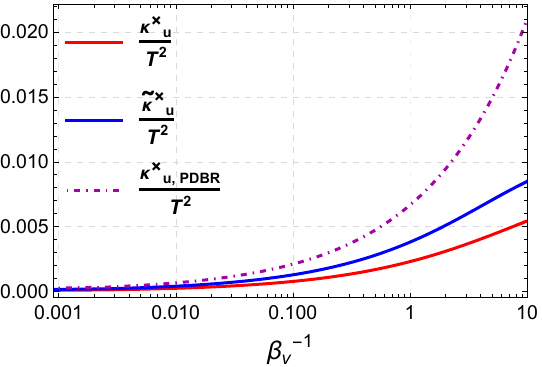}\\
    \includegraphics[width=0.45\textwidth] {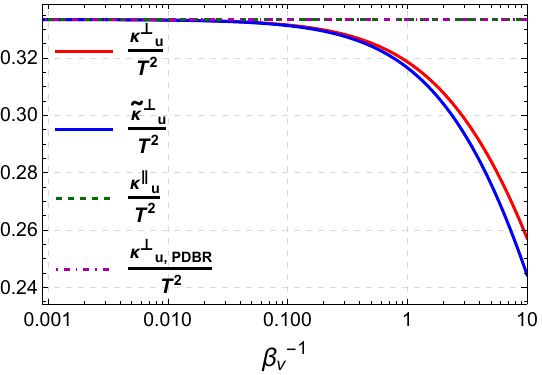} 
    \includegraphics[width=0.48\textwidth] {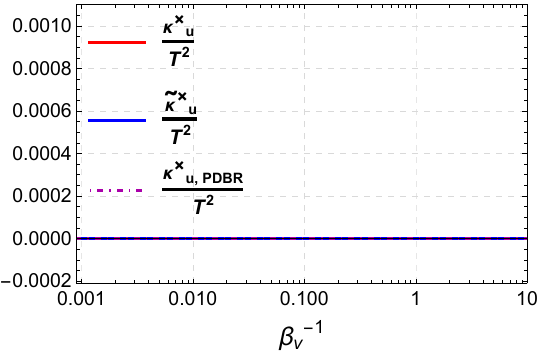}    
    \caption{Perpendicular (left panels) and Hall (right panels) diffusion coefficients for a QGP with a single quark flavor with $Q_u=2\,|e|/3$. The displayed curves refer to the dimensionless chemical potentials $\alpha_u = 0.675$ (upper raw), $\alpha_u = 0.116$ (middle raw) and $\alpha_u=0$ (lower raw), corresponding to the entropy-to-baryon ratios $s/n_B = 50$, $s/n_B = 300$ and $s/n_B\to\infty$, respectively. The dot-dashed lines represent the Navier-Stokes limit of the solution found in Ref.~\cite{Panda:2021pvq} via a Chapman-Enskog approach. While the Hall diffusivity in the PDBR approach looks at least in qualitative agreement with our results up to moderate plasma magnetization, the effect of the magnetic field on the transverse diffusion is completely underestimated.}
    \label{fig:CE}
\end{figure*}
Finally, in Fig.~\ref{fig:CE} we compare our findings obtained through a self-consistent resummation of magnetic-field effects with those presented in Eq.~(40) of Ref.~\cite{Panda:2021pvq} (in the following referred to as PDBR), based on a finite-order truncation of the Chapman-Enskog expansion and referring to the case of a single quark flavor. As shown in the right panels of Fig.~\ref{fig:CE} also the PDBR approach predicts a quenching of the Hall diffusivity/conductivity as the net-quark density decreases. The behavior of $\kappa_{\rm PDBR}^{\,\times}$ as a function of $\beta_V^{\,-1}$ is at least in qualitative agreement with our results up to moderate plasma magnetization. However, this occurrence is directly related to a decrease of the effect of the plasma magnetization on transverse diffusion (left panels), until reaching $\kappa_{\rm PDBR}^{\,\times}=0$ and $\kappa_{\rm PDBR}^{\,\perp}\!=\!\kappa_{\rm PDBR}^{\,\parallel}$  for vanishing quark chemical potential, according to Eq.~(40) of Ref.~\cite{Panda:2021pvq}.
On the contrary, within our approach -- see the discussion after Eq.~(\ref{eq:zero-density}) and~(\ref{eq:alpha0-1f}) -- even in the limit of zero net-quark density and hence in the absence of Hall induced currents one can have a sizable effect of the plasma magnetization on transverse flavor/charge diffusion and conduction, getting
\begin{equation*}
    \widetilde\kappa^{\,\perp}<\kappa^{\,\perp}<\kappa^{\,\parallel}\,.
\end{equation*}

\section{Discussion and perspectives}
In this work, starting from the Boltzmann-Vlasov equation, we have investigated  the diffusion/conduction properties of a strongly-magnetized (meaning that the magnetic pressure can become comparable or larger than the thermal one) Quark Gluon Plasma characterized by multiple quark flavors ($u$, $d$ and $s$) and conserved charges (baryon number ${\cal B}$, electric charge $Q$ and strangeness $S$). The motivation at the basis of our study is that in relativistic HIC's the initial magnetic fields can be huge and in the most peripheral regions of the fireball the energy stored in the magnetic fields can be comparable to the one carried by the plasma particles. Several observables sensitive to the effects of possible large magnetic fields have been proposed in the literature and are at the center of ongoing  active experimental investigation (Chiral Magnetic Effect, $D^0$ vs $\overline D^0$ anisotropic flow, hyperon-antihyperon polarization, etc.). While in the vacuum the initial huge magnetic field generated by the protons of the colliding nuclei would decay by several orders of magnitude during the fireball expansion, if the QGP formed in HIC's behaved as a good electric conductor such a decrease would be much slower. Hence our microscopic calculation of the electric conductivity $\sigma_Q$ of the plasma, found to be connected to other transport coefficients of the medium such as the specific shear-viscosity, is of strong relevance to quantitatively describe the evolution of the magnetic field in the fireball.  At the same time, when the plasma magnetization -- here quantified by the inverse beta-value $\beta_V^{\,-1}$ -- becomes sufficiently large the magnetic field itself can affect the structure of the constitutive equations for the dissipative fields and the associated transport coefficients: this, limited to the equations for the diffusion of net-quark flavors or macroscopic conserved charges, was the core of our study.

Our major results consisted in deriving the non-trivial spacetime tensor structure of the constitutive relations (up to first-order in the expansion parameters ${\rm Kn}$ and $\xi$) for the above diffusion and conduction currents -- with transport coefficients connected by generalized Wiedemann-Franz relations -- arising from the medium spatial anisotropy introduced by the magnetic field. Furthermore, at variance with previous independent studies, we consistently dealt with the presence of multiple quark flavors and conserved charges, making the equations for the respective dissipative currents coupled. In showing our numerical results, we imposed for the macroscopic conserved charges the constraints of relevance to describe the matter produced in HIC's: electric-charge per baryon given by the $Z/A$ ratio of the colliding nuclei, zero net strangeness and a growing entropy-to-baryon ratio $s/n_B$ as one goes from intermediate RHIC center-of-mass energies to top LHC energies.

In our study we also managed to overcome some limitations of previous independent approaches proposed in the literature
\begin{itemize}
\item by fully accounting for the space-time dependence of the quark phase-space distribution and for the fluid velocity and acceleration, at variance with Refs.~\cite{Harutyunyan:2016rxm,Feng:2017tsh,Dash:2020vxk,Dey:2019axu};
\item by performing a self-consistent resummation of the magnetic field effects in the Boltzmann-Vlasov equation, not truncating the corresponding terms at any finite order within a generalized CE expansion, at variance with what done in Refs.~\cite{Panda:2020zhr,Panda:2021pvq}.
\end{itemize}
In particular, this last point was of strong relevance in order to get consistent results for the Hall and transverse conductivities associated to the transport of quark flavor or macroscopic conserved charges.

Our study, far from being complete, can be considered the starting point to achieve a self-consistent description of the effects of a strong plasma magnetization on the dissipative fields, getting for the latter -- as attempted in Refs.~\cite{Hattori:2022hyo,Dey:2019axu,Dash:2020vxk,Kushwah:2024zgd} -- a full set of constitutive relations with a non-trivial tensor structure, involving new transport coefficients with respect to the case of a locally isotropic medium.
Clearly, the subsequent step in order to get a causal set of RMHD equations would consist in going to second order in the present generalized Chapman-Enskog expansion, however still performing a non-perturbative resummation of magnetic effects, in order to address the case of a strongly-magnetized plasma. The final goal would be to arrive to a numerical implementation of the above setup within a full (G)RMHD code. This could be relevant not only for the modeling of electromagnetic effects in high-energy nuclear collisions, but also for applications to astrophysical plasmas. Lastly, possible
polarization and magnetization effects associated to the dipole
moments of the plasma particles, so far neglected, might be included.
We leave all the above important topics to forthcoming studies.

\begin{acknowledgements} 
The authors are grateful to Gabriel S. Denicol, Matthias Kaminski and Akash Jain for useful comments and discussions. F.F. and A.B. acknowledge financial support by MUR within the Prin$\_$2022sm5yas project. L. DZ. acknowledges support from the ICSC — Centro Nazionale di Ricerca in High-Performance Computing; Big Data and Quantum Computing, funded by European Union-NextGenerationEU. 
\end{acknowledgements}

\appendix
\section{Thermodynamic integrals}\label{App:integrals}
In this appendix we provide the reader with more details about the derivation of the flavor-diffusion matrix in a weakly-magnetized plasma. As outlined in Sec.~\ref{sec:strong}, the latter also describes the longitudinal diffusion along the magnetic-field lines in the strong-magnetization case.

The expression of $\kappa_{f\ff}$ in Eq.~(\ref{eq:kappa-weak1}) involves the $A^{\pm}_{f f'}$ matrices defined in Eq.~(\ref{eq:A_ff'}). Taking their difference one obtains
\begin{multline}
A^+_{f\ff}-A^-_{f\ff}=\frac{\tau_R}{\epsilon_p^*}\,\Bigg\{\left[\fpfo\widetilde\fpfo+\fmfo\widetilde\fmfo\right]\,\delta_{f\ff}-\Bigg.\\
\left.-\left[\fpfo\widetilde\fpfo-\fmfo\widetilde\fmfo\right]\,\frac{\epsilon_p^* \,n_\ff}{\varepsilon+P}\label{eq:Ap-Am}
\right\}\,,
\end{multline}
where $\epsilon_p^*\!\equiv\!(-p\!\cdot\!u)$, representing the single-particle energy in the fluid LRF. From Eq.~(\ref{eq:kappa-weak1}) one gets:
\begin{multline}
\kappa_{ff'}=\frac{g_f\,\tau_R}{3}\int d\chi\,\frac{\Delta_{\mu\nu}p^\mu p^\nu}{(-p\!\cdot\!u)}\,\left[\fpfo\widetilde\fpfo+\fmfo\widetilde\fmfo\right]\,\delta_{f\ff}\,-\\
-\frac{g_f\,\tau_R}{3}\int d\chi\,\left( \Delta_{\mu\nu}p^\mu p^\nu\right)\left[\fpfo\widetilde\fpfo-\fmfo\widetilde\fmfo\right]\,\frac{n_\ff}{\varepsilon+P}\,.
\end{multline}
The last expression can be recast in terms of the thermodynamic integrals introduced in Refs.~\cite{Panda:2020zhr,Panda:2021pvq}, transposed here into our different metric signature, namely:
\begin{equation}
\label{eq:kappa_J}    \kappa_{ff'}=g_f\,\tau_R\,J_{21\,,f}^{(1)+}\,\delta_{f\ff}\,-g_f\,\tau_R\,J_{21\,,f}^{(0)-}\,\frac{n_\ff}{\varepsilon+P}\,,
\end{equation}
where we defined the thermodynamic integrals
\begin{align*}
    J_{nq\,,f}^{(m)\pm} \equiv \frac{1}{(2 q + 1)!!} \,\int d\chi\,&(-p\cdot u)^{n-2q-m} \left(\Delta_{\mu \nu} \, p^{\mu} p^{\nu}\right)^q\times\\& \times \left[\fpfo\widetilde\fpfo\pm\fmfo\widetilde\fmfo\right] \,,
\end{align*}
whose definition holds in a generic frame, but which can be conveniently evaluated in the fluid LRF.
For the first integral appearing in Eq.~(\ref{eq:kappa_J}) one gets
\begin{equation}
\label{eq:J1+}
    J_{21\,,f}^{(1)+} = \frac{T^3}{18}\left(1+\frac{3}{\pi^2} \, \alpha_f^2\right)\,.
\end{equation}
In order to express the second term in Eq.~(\ref{eq:kappa_J}) in a more convenient fashion one starts from the kinetic-theory definition of the thermal pressure, isolating in the latter the contribution from flavor $f$:
\begin{equation*}
    P\equiv \frac{1}{3} \, \Delta_{\mu\nu}\, T^{\mu\nu}_{\rm m}=\dots+\frac{g_f}{3}\int d\chi\,\left(\Delta_{\nu\rho}p^\nu p^\rho\right)\left[\fpfo+\fmfo\right]\,.
\end{equation*}
From the latter and the second thermodynamic relation in Eq.~(\ref{eq:therm_rel}), one obtains
\begin{multline}
\label{eq:J0-}
    n_f=\beta\frac{\partial P}{\partial \alpha_f}=\frac{\beta\,g_f}{3}\int d\chi\,\left(\Delta_{\nu\rho}p^\nu p^\rho\right)\,\times\\
    \times\left[\fpfo\widetilde\fpfo-\fmfo\widetilde\fmfo\right] = \beta\,g_f\, J_{21\,,f}^{(0)-}\,\,\,,
    \end{multline}
which can be substituted into the second term in the RHS of Eq.~(\ref{eq:kappa_J}). 

Hence, plugging Eqs.~(\ref{eq:J1+}) and~(\ref{eq:J0-}) into Eq.~(\ref{eq:kappa_J}) one gets the final result
\begin{equation}
    \kappa_{ff'}=\tau_R\left[\frac{g_f T^3}{18}\left(1+\frac{3}{\pi^2}\alpha_{f}^2\right)\delta_{ff'}-\frac{n_f\, n_{f'}T}{\varepsilon+P}\right]\,,
\end{equation}
which coincides with the expression in Eq.~(\ref{eq:kappa-weak}) in the body of the text. 

We end this appendix noting that this last expression is nothing but the generalization of Eq.~(21) in Ref.~\cite{Jaiswal:2015mxa} to the case of a multi-flavored QCD-medium. 
\bibliographystyle{spphys}       
\bibliography{references.bib}   
\end{document}